\documentclass[12pt,letterpaper]{article}

\usepackage[utf8]{inputenc}
\usepackage[T1]{fontenc}
\usepackage{amsmath,amssymb,amsthm,amsfonts,mathtools}
\usepackage[letterpaper,left=1.12in,right=1.12in,top=1.05in,bottom=1.05in]{geometry}
\usepackage{setspace}
\usepackage{needspace}
\usepackage{enumitem}
\usepackage{booktabs,array,tabularx}
\usepackage{xcolor}
\usepackage{hyperref}
\usepackage[round,authoryear]{natbib}
\usepackage[expansion=false]{microtype}
\newcolumntype{Y}{>{\raggedright\arraybackslash}X}

\newif\ifanonymous
\anonymousfalse 
\ifanonymous
\newcommand{\paperauthor}{Anonymous}
\newcommand{\pdfauthorname}{Anonymous}
\else
\newcommand{\paperauthor}{Rui Sun\\ Berkeley Haas \and Junfei Guo\\ Zhejiang University}
\newcommand{\pdfauthorname}{Rui Sun and Junfei Guo}
\fi

\hypersetup{colorlinks=true,linkcolor=black,citecolor=black,urlcolor=black,pdftitle={State-Robust Nash Predictions in Population Games},pdfauthor={\pdfauthorname},pdfkeywords={Population games, Nash equilibrium, reported-state robustness, aggregate misspecification, affine games}}
\theoremstyle{plain}
\newtheorem{theorem}{Theorem}
\newtheorem{proposition}{Proposition}
\newtheorem{observation}{Observation}
\newtheorem{corollary}{Corollary}
\newtheorem{lemma}{Lemma}

\theoremstyle{definition}
\newtheorem{definition}{Definition}
\newtheorem{example}{Example}
\newtheorem{remark}{Remark}

\newcommand{\R}{\mathbb{R}}
\newcommand{\one}{\mathbf{1}}
\newcommand{\eps}{\varepsilon}
\newcommand{\inter}{\operatorname{int}}
\newcommand{\cl}{\operatorname{cl}}
\newcommand{\ri}{\operatorname{ri}}
\newcommand{\coals}{\mathcal{K}}
\newcommand{\CSNB}{\operatorname{CSNB}}
\newcommand{\NE}{\operatorname{NE}}
\newcommand{\SRE}{\operatorname{SRE}}
\newcommand{\BR}{\operatorname{BR}}
\newcommand{\argmax}{\operatorname*{arg\,max}}
\newcommand{\supp}{\operatorname{supp}}

\title{STATE-ROBUST NASH PREDICTIONS IN POPULATION GAMES\thanks{We thank Yuichiro Kamada and Chris Shannon at Berkeley for comments and suggestions.}}
\author{\paperauthor}
\date{}

\begin{document}
\maketitle

\vspace{-0.12in}
\begin{spacing}{1.03}
\small
\noindent\textbf{Abstract.} This paper introduces state-robust equilibrium (SRE), a local validity test for Nash predictions in finite-strategy population games when the payoff-relevant aggregate state may be misspecified. The reported prescription and payoff map are held fixed; only the state used to evaluate payoff comparisons varies. SRE is equivalent to local best-response invariance, absence of structural exposure, and validity along every vanishing interior aggregate-state error. In affine games, tangent-cone, normal-cone, and linear-program tests characterize exposure and identify the exposing population, pure strategy, and aggregate-state direction. The main implication is a sharp negative result: robust mixing requires local payoff identity on the support; in generic affine games SRE reduce to strict pure Nash equilibria, although weak boundary equilibria can survive through feasible-set protection. In affine games with polyhedral local uncertainty regions, the same inequalities yield a deterministic finite diagnostic for reported-state validity.
\end{spacing}

\vspace{0.06in}
\begin{spacing}{1.02}
\small
\noindent\textsc{Keywords:} Population games; Nash equilibrium; reported-state robustness; aggregate misspecification; affine games.

\smallskip
\noindent\textsc{JEL Classification:} C72, C73, C61, D43, O33.
\end{spacing}

\vspace{0.12in}


\section{Introduction}\label{sec:introduction}

Population games make aggregate behavior payoff-relevant. A state records, for each population, how much mass is assigned to each strategy; payoffs are evaluated as functions of that distribution. This representation is central in congestion, matching, platform competition, technology adoption, and network-externality environments. It also creates a basic robustness problem. The aggregate state used in payoff comparisons may be reported by an institution, estimated from data, sampled from a population, forecast by a platform, or implemented with error. A Nash prediction can therefore be correct at the reported state and fail at nearby states that the analyst cannot rule out.

This paper studies the following reported-state experiment. Fix a payoff map $F$ and a reported aggregate prescription $x$: population $p$ is predicted or recommended to use distribution $x_p$. For each nearby feasible aggregate statistic $y$, evaluate the payoff from continuing to use $x_p$ and from deviating to any $z_p\in X_p$ at that same evaluation state $y$. The report passes the test if $x_p$ remains a best response for every population throughout a relative neighborhood of $x$. The nearby $y$ is not a new prediction, not a tremble, not a payoff perturbation, and not a consistency condition for realized play. It is the payoff-relevant aggregate used to evaluate the fixed report.

A state $x$ is a state-robust equilibrium, or SRE,\footnote{The acronym denotes state-robust equilibrium in this paper. \citet{khan2012randommatrix} use the same phrase for a state-independent strategy profile that is an equilibrium for every realized payoff state in a random matrix game. Here the payoff map and reported prescription are fixed, and robustness is local in the payoff-evaluation aggregate state. The concept is also distinct from strategically robust equilibrium in \citet{lanzetti2025strategically}.} if
\begin{equation}\label{eq:intro-localbr}
\exists V\ni x\text{ such that }x_p\in\BR_p(y)\quad \forall p\in P,\ \forall y\in V\cap\inter(X),
\end{equation}
where $V$ is a relative neighborhood in the product simplex. Thus SRE is a local validity test for the best-response content of a Nash prediction. It is not an existence theorem and it may be empty. When it is empty, the conclusion is informative: no static Nash prediction in the game is locally valid under arbitrarily small aggregate-state misspecification without additional structure.

Hawk--Dove gives the simplest illustration. At the mixed Nash state, Hawk and Dove payoffs are equal. A slightly higher Hawk share makes Dove strictly profitable relative to the candidate mixture; a slightly lower Hawk share makes Hawk strictly profitable. The mixed equilibrium is therefore protected only by equality at one point. A strict pure equilibrium, by contrast, remains a best response throughout a neighborhood. SRE formalizes this distinction without specifying a mutation process, a learning dynamic, a tremble distribution, or a payoff perturbation.

The equivalent negative notion is structural exposure. A pure strategy exposes a candidate state when its payoff gap against the fixed report becomes strictly positive along interior states converging to that candidate. Although exposure is written as a sequence, it is a local open-set property: the candidate lies in the closure of a strict-improvement region. This formulation separates the proposal from the evaluation state. The proposal is the alternative distribution being tested; the perturbation is only the aggregate statistic at which payoffs are evaluated.

Before defining SRE, the paper records a benchmark about coalitional language. If coalition proposals are evaluated at the candidate state, one member's reallocation does not enter another member's payoff comparison. Candidate-state no-blocking therefore coincides with Nash equilibrium. The observation is intentionally elementary: under this payoff convention the benchmark is Nash; the substantive restriction is the subsequent local reported-state test.

The first main result gives an exact characterization. For continuous finite-strategy population games, the following are equivalent: local best-response invariance as in \eqref{eq:intro-localbr}; absence of structural exposure; eventual validity along every vanishing sequence of interior aggregate-state errors; and absence of nearby pure strict-improvement regions. Finiteness is essential. Once every pure strict improvement is ruled out locally, finitely many neighborhoods can be intersected to obtain a single neighborhood on which all best-response inequalities hold. A maximality corollary gives a representation result: among all prediction correspondences required to remain valid under arbitrary vanishing aggregate-state error, SRE is the largest.

For affine games, the test becomes geometric and finite. For population $p$ and pure strategy $i$, define the payoff gap
\[
 h_{p,i}(y;x)=m_pF_{p,i}(y)-x_p\cdot F_p(y),
\]
where $x$ is the report and $y$ is the evaluation state. At a Nash state, only zero gaps can expose. A zero gap is exposed exactly when an inward tangent direction raises it strictly. Equivalently, the gap is not exposed exactly when its derivative belongs to the normal cone of the product simplex at the candidate. Hence SRE membership in affine games can be checked by at most $\sum_{p\in P}n_p$ linear programs. When a candidate fails, the program identifies the exposing population, pure strategy, and aggregate-state direction.

This geometry yields an exposed-indifference decomposition of the Nash set. The difference $\NE(\Gamma)\setminus\SRE(\Gamma)$ is precisely the finite union of Nash states at which some zero payoff gap is exposed by an inward feasible perturbation. Boundary states are treated endogenously. A weak boundary equilibrium may survive because the direction that would break the tie is infeasible; another weak equilibrium may fail because a feasible inward direction breaks the tie. Thus SRE removes locally exposed equality, not all weak equilibria.

The sharpest implication concerns mixing. If a state is an SRE, then any two pure strategies used with positive mass by the same population must have identical payoff functions on a neighborhood of the candidate. In affine games this local identity extends to all of $X$. Consequently, outside a finite union of proper payoff-identity subspaces, every affine SRE is pure; after excluding payoff-tie hyperplanes at pure states, SRE coincides with the set of strict pure Nash equilibria. This generic collapse is the paper's negative theorem, not a definition of the concept. Absent genuine payoff identities or feasible-set protection at the boundary, weak and mixed Nash predictions are locally exposed to arbitrarily small aggregate-state misspecification. SRE therefore identifies exactly what additional structure is needed for non-strict predictions to survive the reported-state test.

The finite test also gives an operational diagnostic. If a reported state is accompanied by a local region of possible payoff-evaluation states, the analyst can ask whether all candidate best-response inequalities hold throughout that region. In affine games with closed polyhedral regions, this is again a finite collection of linear programs. A shrinking-region result shows that regular local uncertainty regions--sets satisfying explicit inner-ball and outer-shrinking conditions--eventually accept exactly the SRE states and reject exactly the exposed states. The step is deterministic; statistical confidence regions can be used only when an external statistical procedure supplies the required shrinking, inner-ball, and coverage properties.

The examples illustrate the scope of the concept. Hawk--Dove and rock--paper--scissors have exposed mixed equilibria and empty SRE sets; coordination, platform adoption, and standards adoption have robust pure anchors. Boundary and payoff-identity examples show that weak equilibria and robust mixing can survive when protected by feasibility or genuine local payoff equivalence. SRE does not rank strict local anchors; welfare, history, risk dominance, learning, institutions, or global-game arguments may still be needed for that task.

The paper is related to several literatures on robustness and refinement, but the object is different. Perfection, sequential equilibrium, strategic stability, persistence, and essential equilibrium rely on perturbations of strategies, beliefs, assessments, games, or strategic forms \citep{selten1975reexamination,kreps1982sequential,kohlberg1986strategic,vandamme1991stability,wujiang1962essential,kalai1984persistent,fudenberg1988robustness}. Robust predictions and robustness to incomplete information vary higher-order beliefs or information structures \citep{kajii1997robustness,bergemann2013robust,morris2005generalized,bergemann2016bayes}. Large-game, continuous-game, and robust aggregative concepts impose different robustness, ambiguity, payoff-perturbation, or admissibility requirements \citep{kalai2004large,chen2022robust,lotidis2025robust,lanzetti2025strategically,feik2026strategically}. Evolutionary stability, local superiority, neighborhood-invader strategies, and static stability compare resident and mutant or neighboring population states \citep{maynardsmith1973logic,sandholm2010population,apaloo1997neighborhood,milchtaich2020static}. SRE instead fixes the payoff map and the reported prescription and varies only the aggregate state used to evaluate that prescription. Section~\ref{subsec:otherrefinements} gives the detailed comparison.

The rest of the paper is organized as follows. Section~\ref{sec:model} introduces the primitives and the candidate-state no-blocking benchmark. Section~\ref{sec:sre} defines SRE and structural exposure and proves the exact characterization. Section~\ref{sec:affine} develops the tangent-cone, normal-cone, linear-program, exposed-indifference, support-equality, generic-purity, and codimension results. Section~\ref{sec:applications} gives examples and applications. Section~\ref{sec:discussion} compares SRE with related concepts and develops the uncertainty-region diagnostic. Section~\ref{sec:conclusion} concludes. Proofs are in Appendix~\ref{sec:proofs}.


\section{Population Games}\label{sec:model}

\subsection{Primitives}\label{subsec:primitives}

A finite-strategy population game is a tuple
\[
\Gamma=(P,(m_p,S_p,X_p,F_p)_{p\in P}),
\]
where $P=\{1,\ldots,P_0\}$ is a finite set of populations. For each population $p$, $m_p>0$ is its mass, $S_p=\{1,\ldots,n_p\}$ is its finite strategy set, and
\[
X_p=\left\{x_p\in\R_+^{n_p}:\sum_{i\in S_p}x_{p,i}=m_p\right\}
\]
is the simplex of population-$p$ distributions. The joint state space is
\[
X=\prod_{p\in P}X_p\subset\R^n,\qquad n=\sum_{p\in P}n_p.
\]
The payoff function $F:X\to\R^n$ is continuous. Its component $F_{p,i}(x)$ is the payoff to strategy $i\in S_p$ for members of population $p$ at state $x$, and $F_p(x)=(F_{p,i}(x))_{i\in S_p}$.

For any distribution $z_p\in X_p$, the aggregate payoff earned by population $p$ when it uses $z_p$ and payoffs are evaluated at state $y$ is
\[
U_p(z_p,y)=z_p\cdot F_p(y).
\]
The per-capita payoff is $m_p^{-1}U_p(z_p,y)$. Since $m_p$ is fixed for a population, all payoff comparisons are unchanged by using aggregate rather than per-capita payoffs.

The best-response correspondence of population $p$ at state $y$ is
\[
\BR_p(y)=\argmax_{z_p\in X_p} z_p\cdot F_p(y).
\]
A state $x$ is a Nash equilibrium if $x_p\in\BR_p(x)$ for every $p\in P$. The set of Nash equilibria is denoted $\NE(\Gamma)$.

For a pure strategy $i\in S_p$, let $e_{p,i}\in X_p$ denote the pure distribution assigning all mass $m_p$ to $i$. Thus $e_{p,i}\cdot F_p(y)=m_pF_{p,i}(y)$. For a pure profile $\sigma=(\sigma_p)_{p\in P}\in\prod_{p\in P}S_p$, write $e^\sigma=(e_{p,\sigma_p})_{p\in P}$. A pure state $e^\sigma$ is a strict pure Nash equilibrium if, for every population $p\in P$ and every $j\in S_p\setminus\{\sigma_p\}$,
\[
F_{p,\sigma_p}(e^\sigma)>F_{p,j}(e^\sigma).
\]
Throughout, $\inter(X)$ denotes relative interior in the affine hull of $X$. All neighborhoods are relative to $X$; whenever a neighborhood appears in a local condition, it can be replaced by a relatively open smaller neighborhood.

The tangent cone to $X$ at $x$ is
\[
T_xX=\{d\in\R^n:\exists t_k\downarrow0\text{ such that }x+t_kd\in X\},
\]
and its relative interior is $\ri(T_xX)$. The normal cone is the polar cone
\[
N_X(x)=\{z\in\R^n:z\cdot d\le0\text{ for all }d\in T_xX\}.
\]
Because $X$ is a product of finite simplices, these cones are polyhedral.

\subsection{Candidate-state no-blocking as a benchmark observation}\label{subsec:benchmark}

Let $\coals=\{S\subseteq P:S\ne\varnothing\}$ be the set of non-empty coalitions of populations. A proposal by coalition $S$ is a vector $z_S=(z_p)_{p\in S}\in\prod_{p\in S}X_p$.

\begin{definition}[Candidate-state block]\label{def:candidate-block}
A coalition $S\in\coals$ candidate-state blocks $x\in X$ at proposal $z_S$ if
\begin{equation}\label{eq:candidate-block}
z_p\cdot F_p(x)>x_p\cdot F_p(x),\qquad \forall p\in S.
\end{equation}
The candidate-state no-blocking set is
\[
\CSNB(\Gamma)=\{x\in X:\text{no coalition candidate-state blocks }x\}.
\]
\end{definition}

The notation $\CSNB$ avoids confusion with correlated equilibrium. The construction recalls the core and $\alpha$-core tradition in cooperative game theory \citep{aumann1961core,scarf1971existence,kajii1992generalization}, but the payoff convention is narrower: coalition proposals are evaluated at the candidate state rather than at a post-deviation state. The observation records the benchmark implied by that convention, not a substantive cooperative-core theorem.

\begin{observation}[Candidate-state no-blocking equals Nash]\label{obs:csnb-ne}
For every finite-strategy population game with continuous payoffs,
\[
\CSNB(\Gamma)=\NE(\Gamma).
\]
\end{observation}

The argument is immediate, and that is precisely the point. Since the payoff of each coalition member is evaluated at the candidate state $x$, another member's proposal has no effect on that member's inequality in~\eqref{eq:candidate-block}. A non-singleton block exists only if each of its members already has a singleton improvement. Conversely, any failure of the best-response condition is a singleton block. Thus candidate-state blocking is not an independent cooperative refinement; under this payoff convention, the benchmark is exactly Nash equilibrium.

\begin{remark}[Genuine coalitional externalities]\label{rem:postdev}
One can define a genuinely coalitional theory by evaluating payoffs at the post-deviation state $(z_S,x_{-S})$, or at a nearby post-deviation perturbation. That alternative is not the object of this paper. The present paper deliberately keeps the candidate-state payoff convention. Observation~\ref{obs:csnb-ne} then identifies the relevant noncooperative benchmark before the local reported-state robustness test is applied.
\end{remark}


\section{Structural Exposure and State-Robust Equilibrium}\label{sec:sre}

\subsection{Definitions}\label{subsec:definitions}

The definition below separates two uses of the same aggregate coordinates. The reported state $x$ is the prescription whose optimality is assessed. The nearby state $y$ is the payoff-relevant aggregate statistic at which payoff comparisons are evaluated. The same state space is used for both objects because measurement error, sampling error, and implementation error are naturally expressed in population-share coordinates. The model does not require $y$ to be the realized distribution generated by agents following $x$, nor does it require $y$ to be an equilibrium state.

\begin{definition}[Perturbation]\label{def:perturbation}
A perturbation of $x\in X$ is a sequence $(y^k)_{k\ge1}$ such that $y^k\in\inter(X)$ for every $k$ and $y^k\to x$.
\end{definition}

Perturbations exist for every $x\in X$: fix any $\omega\in\inter(X)$ and set $y^k=(1-1/k)x+(1/k)\omega$.

\begin{definition}[State-robust equilibrium]\label{def:sre}
A state $x\in X$ is a state-robust equilibrium, or SRE, if there exists a relative neighborhood $V$ of $x$ in $X$ such that
\begin{equation}\label{eq:localbr-definition}
 x_p\in\BR_p(y),\quad \forall p\in P,\ \forall y\in V\cap\inter(X).
\end{equation}
The set of state-robust equilibria is denoted $\SRE(\Gamma)$.
\end{definition}

The definition keeps the reported prescription and payoff map fixed and varies only the payoff-evaluation state. The payoff comparison is between a proposed distribution $z_p$ and the reported distribution $x_p$, both evaluated at $y$. The interior restriction makes the local state-error test evaluate the report along full-support aggregate states. Deviations themselves remain unrestricted in $X_p$; the restriction is on the payoff-evaluation state, not on the feasible deviation set.

\begin{definition}[Structural exposure]\label{def:exposure}
A coalition $S\in\coals$ structurally exposes $x\in X$ at proposal $z_S\in\prod_{p\in S}X_p$ if there exists a perturbation $(y^k)$ of $x$ such that, for every sufficiently large $k$,
\begin{equation}\label{eq:exposure}
 z_p\cdot F_p(y^k)>x_p\cdot F_p(y^k),\quad \forall p\in S.
\end{equation}
A state is structurally exposed if some coalition structurally exposes it.
\end{definition}

Although exposure is witnessed by a sequence, it is not a zero-measure path objection. For each fixed proposal, the strict-improvement region is relatively open in $X$. A witnessing path therefore means that every neighborhood of the candidate contains an open set of interior states at which the proposal is strictly profitable. The equivalence theorem below shows that SRE is exactly the absence of such local strict improvement.

\subsection{Singleton and pure-deviation reductions}\label{subsec:reductions}

The candidate-state separability from Observation~\ref{obs:csnb-ne} persists under structural exposure.

\begin{lemma}[Singleton reduction]\label{lem:singleton}
A state $x$ is structurally exposed by some coalition if and only if it is structurally exposed by some singleton population.
\end{lemma}

Therefore the relevant deviations are population-level reallocations. For $p\in P$, $z_p\in X_p$, and $x\in X$, define the strict-improvement region
\begin{equation}\label{eq:Up}
\mathcal U_p(z_p;x)=\{y\in\inter(X):z_p\cdot F_p(y)>x_p\cdot F_p(y)\}.
\end{equation}

\begin{proposition}[Closure characterization]\label{prop:closure}
For fixed $p$, $z_p$, and $x$, the proposal $z_p$ structurally exposes $x$ if and only if
\[
x\in\cl\mathcal U_p(z_p;x),
\]
where the closure is taken in $X$.
\end{proposition}

Since $z_p\mapsto z_p\cdot F_p(y)$ is linear on the simplex $X_p$, mixed proposals can be replaced by pure proposals.

\begin{proposition}[Pure-deviation reduction]\label{prop:pure}
A state $x$ is structurally exposed if and only if there exist a population $p\in P$ and a pure strategy $i\in S_p$ such that $e_{p,i}$ structurally exposes $x$.
\end{proposition}

For pure deviations write
\begin{equation}\label{eq:gap}
h_{p,i}(y;x)=e_{p,i}\cdot F_p(y)-x_p\cdot F_p(y)=m_pF_{p,i}(y)-x_p\cdot F_p(y).
\end{equation}
Then $e_{p,i}$ structurally exposes $x$ if and only if
\[
x\in\cl\{y\in\inter(X):h_{p,i}(y;x)>0\}.
\]

\subsection{Exact local-best-response characterization}\label{subsec:localbr}

\begin{theorem}[Local best-response characterization]\label{thm:localbr}
For every continuous finite-strategy population game and every $x\in X$, the following statements are equivalent:
\begin{enumerate}[label=\textup{(\roman*)},leftmargin=0.38in]
\item $x\in\SRE(\Gamma)$;
\item $x$ is not structurally exposed;
\item for every perturbation $(y^k)$ of $x$, there is $K$ such that
\[
 x_p\in\BR_p(y^k),\quad \forall p\in P,\ \forall k\ge K;
\]
\item for every $p\in P$ and every $i\in S_p$,
\[
x\notin\cl\{y\in\inter(X):h_{p,i}(y;x)>0\}.
\]
\end{enumerate}
Consequently, $\SRE(\Gamma)\subseteq\NE(\Gamma)$.
\end{theorem}

The theorem gives the economic meaning of the refinement. A Nash equilibrium is state-robust precisely when the candidate distributions continue to be best responses after all sufficiently small interior changes in the aggregate state. Item (iii) is the vanishing-state-error formulation: no matter how the true aggregate state approaches the reported candidate through interior states, the reported population distributions eventually remain optimal. If this neighborhood does not exist, then some pure strategy becomes strictly better arbitrarily close to the candidate, and the equilibrium is structurally exposed.

The finite-strategy assumption is doing useful work in this theorem. For each pure strategy, absence of exposure gives a neighborhood on which that pure strategy does not strictly improve on the candidate. Since there are finitely many pure strategies, these neighborhoods can be intersected. The theorem is not a compactness-free statement for arbitrary strategy spaces: with infinitely many strategies, one would need additional compactness and upper-hemicontinuity conditions strong enough to prevent improving strategies from escaping along a sequence.

Because $\inter(X)$ is dense in $X$ and the payoff maps are continuous, item (i) is equivalent to the same local best-response condition on a possibly smaller relative neighborhood $V$ without the restriction $y\in\inter(X)$. The interior convention asks whether the fixed report remains valid along full-support evaluation states approaching the candidate, while the set of deviations remains the whole simplex $X_p$. The resulting local best-response characterization is not weakened by this convention.

The theorem also separates SRE from ordinary upper hemicontinuity of best responses. It does not ask whether some best response near $x$ is close to $x_p$. It asks whether the fixed candidate distribution $x_p$ itself remains a best response at nearby states. This is a stronger local invariance property and is exactly the property that fails at mixed anti-coordination equilibria.

\subsection{A maximality representation}\label{subsec:maximality}

The preceding theorem can be restated as a maximality representation for the reported-state robustness requirement. Let a solution correspondence $R$ assign to each finite-strategy population game $\Gamma$ a subset $R(\Gamma)\subseteq X$.

\begin{definition}[Vanishing-state-error validity]\label{def:validity}
A solution correspondence $R$ is valid under vanishing aggregate-state error if, whenever $x\in R(\Gamma)$ and $(y^k)$ is any perturbation of $x$, there exists $K$ such that
\[
x_p\in\BR_p(y^k),\quad \forall p\in P,\ \forall k\ge K.
\]
\end{definition}

This requirement uses only primitive best responses and the idea that the reported aggregate state may be an approximation to nearby true aggregate states. It does not mention structural exposure. It says that a retained reported prediction must remain correct under every sufficiently small interior measurement or implementation error.

\begin{corollary}[Maximality under vanishing aggregate-state error]\label{cor:maximality}
On the class of continuous finite-strategy population games, SRE is the largest solution correspondence valid under vanishing aggregate-state error. That is, if $R$ is any correspondence satisfying Definition~\ref{def:validity}, then
\[
R(\Gamma)\subseteq\SRE(\Gamma)\quad\text{for every }\Gamma,
\]
and the correspondence $\Gamma\mapsto\SRE(\Gamma)$ itself satisfies Definition~\ref{def:validity}.
\end{corollary}

The corollary is a representation result, not a claim that every application must impose the reported-state requirement. Once that requirement is imposed, however, the conclusion is exact: any prediction rule valid under arbitrary vanishing aggregate-state error must be a subset of SRE, and SRE discards nothing beyond what the requirement forces.

A strict pure Nash equilibrium automatically satisfies Theorem~\ref{thm:localbr}'s local condition.

\begin{corollary}[Strict pure equilibria survive]\label{cor:strictpure}
Every strict pure Nash equilibrium is a state-robust equilibrium.
\end{corollary}

The converse is false in degenerate games. If all strategies used by a population have the same payoff on a neighborhood, then mixtures can be state-robust. The support-equality theorem below pinpoints this degeneracy: a mixed SRE requires local payoff equality among all strategies in its support, and in affine games this local equality becomes global equality on $X$.

\begin{remark}[Closure and stronger robustness tests]\label{rem:taxonomy}
The closure test is the weakest local strict-improvement test. It declares a candidate exposed when a strict-improvement region comes arbitrarily close to the candidate. Since payoffs are continuous, each strict-improvement region is relatively open in $X$; hence exposure is equivalently the statement that every neighborhood of the candidate contains an open set of interior states at which the fixed proposal is profitable. One could impose stronger tests, for example by requiring profitability for almost every local perturbation or for every sufficiently small interior perturbation in a neighborhood. The present paper uses the closure test because it is the minimal local way to rule out equilibria protected only by pointwise equality. In affine games, the closure test is also the natural geometric object: strict-improvement regions are intersections of the simplex with open half-spaces, and exposure is exactly the existence of a favorable tangent direction.

State-robustness does not claim that an exposed state is unstable under every perturbation. It claims that the candidate's optimality is not locally insulated from strict improvement. By Theorem~\ref{thm:localbr}, SRE is necessary and sufficient for local best-response robustness of the fixed candidate distribution with respect to interior aggregate-state perturbations. It is not, and is not meant to be, a sufficient condition for dynamic stability, welfare dominance, stochastic stability, robustness to payoff perturbations, or selection among multiple strict local anchors.
\end{remark}


\section{Affine Games: Geometry and Generic Exposure}\label{sec:affine}

\subsection{Tangent-cone test}\label{subsec:tangent}

A population game is affine if each payoff component is affine on $X$. Equivalently, for each population $p$ there exist a matrix $A_p$ and a vector $b_p$ such that
\[
F_p(y)=A_py+b_p.
\]
For fixed $x$, the pure gap $h_{p,i}(\cdot;x)$ is affine. Let
\[
L=\operatorname{span}(X-X)
\]
be the tangent space of the affine hull of $X$. Affine functions on $X$ may have more than one ambient representation in $\R^n$. Accordingly, $D_y h_{p,i}(x;x)$ denotes the induced linear functional on $L$. When it is written as an ambient vector, any representative of that functional may be used.

\begin{lemma}[Interior tangent directions]\label{lem:interior-tangent}
For every $x\in X$, the following statements hold.
\begin{enumerate}[label=(\roman*),leftmargin=0.28in]
\item If $y\in\inter(X)$, then $y-x\in\ri(T_xX)$.
\item If $d\in\ri(T_xX)$, then $x+td\in\inter(X)$ for all sufficiently small $t>0$.
\end{enumerate}
\end{lemma}

\begin{lemma}[Representative invariance]\label{lem:representative}
Let $a$ and $a'$ be two ambient vectors representing the same affine derivative on $X$. Then $a\cdot d=a'\cdot d$ for every $d\in T_xX$ and every $x\in X$. Consequently, the tangent inequalities in Theorem~\ref{thm:tangent} and the normal-cone condition $a\in N_X(x)$ in Proposition~\ref{prop:normal} are independent of the chosen ambient affine representation.
\end{lemma}

\begin{theorem}[Affine tangent-cone characterization]\label{thm:tangent}
Suppose $\Gamma$ is affine. Fix $x\in X$, $p\in P$, and $i\in S_p$. The pure deviation $e_{p,i}$ structurally exposes $x$ if and only if one of the following two conditions holds:
\begin{enumerate}[label=(\roman*),leftmargin=0.28in]
\item $h_{p,i}(x;x)>0$;
\item $h_{p,i}(x;x)=0$ and there exists $d\in\ri(T_xX)$ such that
\begin{equation}\label{eq:tangent-ineq}
D_y h_{p,i}(x;x)[d]>0.
\end{equation}
\end{enumerate}
If $h_{p,i}(x;x)<0$, the pure deviation does not structurally expose $x$.
\end{theorem}

At a Nash state, $h_{p,i}(x;x)\le0$ for every $p,i$. Hence the only affine exposures that matter on the equilibrium set are zero-gap exposures: a pure strategy tied with the candidate at $x$ becomes strictly better in an inward tangent direction.

\subsection{Normal-cone certificate and finite test}\label{subsec:normal}

The tangent test has an equivalent normal-cone form. The argument uses standard finite-dimensional cone duality \citep{rockafellar1970convex,aliprantis2006infinite}.

\begin{proposition}[Normal-cone certificate]\label{prop:normal}
Suppose $\Gamma$ is affine and $h_{p,i}(x;x)=0$. The pure deviation $e_{p,i}$ does not structurally expose $x$ if and only if
\begin{equation}\label{eq:normal-cert}
D_y h_{p,i}(x;x)\in N_X(x).
\end{equation}
\end{proposition}

For implementation, define
\begin{equation}\label{eq:lp}
\Psi_{p,i}(x)=\max\left\{D_y h_{p,i}(x;x)[d]:d\in T_xX,\ \|d\|_\infty\le1\right\}.
\end{equation}
The optimization problem is a linear program because $T_xX$ is polyhedral and $h_{p,i}$ is affine. Since $d=0$ is feasible, $\Psi_{p,i}(x)\ge0$.

When $F_p(y)=A_py+b_p$, a convenient ambient representative of the derivative of the pure gap is
\begin{equation}\label{eq:affine-derivative}
a_{p,i}(x)=m_pA_{p,i\cdot}-x_p^\top A_p,
\end{equation}
where $A_{p,i\cdot}$ is the $i$th row of $A_p$. Thus \eqref{eq:lp} can be implemented as
\[
\begin{array}{ll}
\text{maximize} & a_{p,i}(x)\cdot d \\
\text{subject to} & \displaystyle\sum_{j\in S_q}d_{q,j}=0,\qquad q\in P, \\
& d_{q,j}\ge0\quad\text{whenever }x_{q,j}=0, \\
& -1\le d_{q,j}\le1,\qquad q\in P,\ j\in S_q.
\end{array}
\]
The last line is only the normalization $\|d\|_\infty\le1$ written coordinate by coordinate. Only the induced functional on $\operatorname{span}(X-X)$ matters, so the value of the program is unchanged by replacing \eqref{eq:affine-derivative} with any equivalent affine representation.

\begin{corollary}[Finite linear-program test]\label{cor:lp}
Suppose $\Gamma$ is affine. Then $x\in\SRE(\Gamma)$ if and only if, for every $p\in P$ and $i\in S_p$, either
\[
h_{p,i}(x;x)<0,
\]
or
\[
h_{p,i}(x;x)=0\quad\text{and}\quad \Psi_{p,i}(x)=0.
\]
If $h_{p,i}(x;x)>0$ for some pair $(p,i)$, membership fails immediately because $x$ is not Nash. Thus membership in $\SRE(\Gamma)$ is checked by at most $\sum_{p\in P}n_p$ linear programs.
\end{corollary}

The explicit tangent and normal cones for product simplices are useful. For population $q$,
\[
T_{x_q}X_q=\left\{d_q\in\R^{n_q}:\sum_{j\in S_q}d_{q,j}=0,\ d_{q,j}\ge0\text{ whenever }x_{q,j}=0\right\}.
\]
The relative interior replaces the weak inequalities by strict inequalities at inactive strategies. The normal cone is
\[
N_{X_q}(x_q)=\left\{z_q:\exists a_q\in\R\text{ such that }z_{q,j}=a_q\text{ on }\supp(x_q),\ z_{q,j}\le a_q\text{ off }\supp(x_q)\right\}.
\]
Since $X$ is a product, $T_xX=\prod_qT_{x_q}X_q$ and $N_X(x)=\prod_qN_{X_q}(x_q)$.

At an interior state, $T_xX$ is the affine tangent space of the product simplex, so exposure asks whether the relevant payoff-gradient vector is nonzero on that tangent space. At a boundary state, the tangent cone is one-sided. This distinction is economically meaningful. Boundary equilibria can be robust not because nearby payoff gaps are flat, but because the directions that would make a deviation profitable are infeasible: they would require moving negative mass out of a strategy that is already absent. The normal-cone certificate records exactly this boundary protection.

The linear program in \eqref{eq:lp} is normalized by $\|d\|_\infty\le1$ only to avoid scale indeterminacy. Because the inequalities are homogeneous in the direction and $d=0$ is feasible, $\Psi_{p,i}(x)>0$ is equivalent to the existence of an exposing tangent direction, while $\Psi_{p,i}(x)=0$ means that every feasible first-order movement weakly decreases the zero gap or leaves it flat. If an optimizer with positive value lies on the boundary of $T_xX$, any sufficiently small convex combination with a fixed vector in $\ri(T_xX)$ still has positive directional derivative and generates an interior perturbation path. In affine games this is exactly local protection against that pure deviation.

\subsection{Exposed-indifference decomposition}\label{subsec:knife}

For an affine game define the exposed indifference set associated with population $p$ and strategy $i$ by
\begin{equation}\label{eq:Kpi}
K_{p,i}(\Gamma)=\bigl\{\,x\in\NE(\Gamma):h_{p,i}(x;x)=0\text{ and }\exists d\in\ri(T_xX)\text{ with }D_y h_{p,i}(x;x)[d]>0\,\bigr\}.
\end{equation}

\begin{theorem}[Exposed-indifference decomposition]\label{thm:knife}
For every affine finite-strategy population game,
\begin{equation}\label{eq:knife}
\NE(\Gamma)\setminus\SRE(\Gamma)=\bigcup_{p\in P}\ \bigcup_{i\in S_p}K_{p,i}(\Gamma).
\end{equation}
\end{theorem}

This theorem is the main geometric statement. State-robustness subtracts exactly the Nash states at which a zero payoff gap is locally exposed by an inward state perturbation. Non-existence of SRE is therefore informative: it means that every Nash prediction is supported only by locally exposed indifference.

\subsection{Support equality and generic restrictions}\label{subsec:generic}

The local best-response characterization implies a restriction that is stronger than the full-support case. A mixed state can be robust only if every pure strategy used with positive mass remains payoff-equivalent to every other used pure strategy under nearby state perturbations.

\Needspace{9\baselineskip}
\begin{theorem}[Support equality]\label{thm:support-equality}
For every continuous finite-strategy population game, if $x\in\SRE(\Gamma)$, then there exists a relative neighborhood $V$ of $x$ in $X$ such that, for every population $p$ and every pair $i,j\in\supp(x_p)$,
\begin{equation}\label{eq:support-equality-local}
F_{p,i}(y)=F_{p,j}(y)\qquad \forall y\in V.
\end{equation}
If $\Gamma$ is affine, then the equality is global:
\begin{equation}\label{eq:support-equality-global}
F_{p,i}(y)=F_{p,j}(y)\qquad \forall y\in X,
\end{equation}
for every $p$ and every $i,j\in\supp(x_p)$.
\end{theorem}

The theorem says that a mixed robust candidate is not merely a pointwise indifference. It requires a local payoff identity on the support. In affine games, local identity on a relative open subset forces identity on the entire simplex.

Let $\mathcal A_X$ denote the vector space of affine real-valued functions on $X$, and treat affine payoff maps as elements of the quotient space $\prod_{p\in P}\mathcal A_X^{n_p}$, so two coefficient representations that induce the same affine functions on $X$ are identified. For a pure profile $\sigma=(\sigma_p)_{p\in P}\in\prod_{p\in P}S_p$, write $e^\sigma=(e_{p,\sigma_p})_{p\in P}$. Define the payoff-identity subspaces
\[
\mathcal I=\bigcup_{p\in P}\ \bigcup_{\substack{i,j\in S_p\\ i<j}}
\{\Gamma:F_{p,i}=F_{p,j}\text{ on }X\}
\]
and the pure-state payoff-tie hyperplanes
\[
\mathcal H=\bigcup_{\sigma\in\prod_{q\in P}S_q}\ \bigcup_{p\in P}\ \bigcup_{j\in S_p\setminus\{\sigma_p\}}
\{\Gamma:F_{p,\sigma_p}(e^\sigma)=F_{p,j}(e^\sigma)\}.
\]
Here ``generic'' means outside such a finite union of proper linear subspaces or hyperplanes in the finite-dimensional affine payoff space; after any choice of coordinates, the exceptional set is closed, nowhere dense, and Lebesgue null.

\begin{corollary}[Generic purity in affine games]\label{cor:generic-purity}
In the affine payoff space $\prod_{p\in P}\mathcal A_X^{n_p}$, every game outside $\mathcal I$ has only pure state-robust equilibria. Every game outside $\mathcal I\cup\mathcal H$ satisfies
\[
\SRE(\Gamma)=\{\text{strict pure Nash equilibria of }\Gamma\}.
\]
\end{corollary}

Corollary~\ref{cor:generic-purity} should be read as a sharp implication of the reported-state test, not as a definition of SRE by strictness. Outside $\mathcal H$, every pure Nash equilibrium is strict because no unchosen pure strategy can tie the chosen one at the pure profile. The exceptional payoff-identity subspaces need not be economically irrelevant: regulated prices, technological standards, capacity constraints, institutional rules, or common-value components can make local payoff identities substantive rather than accidental. Outside such structures, mixed Nash predictions are generically exposed because their support indifference is only pointwise. Boundary protection remains a separate force, captured by the normal cone.

A simple identity-based case illustrates why the exceptional set can be substantive. Suppose a one-population environment has two institutionally equivalent compliance modes, $1$ and $2$, with $F_1(y)=F_2(y)$ throughout a neighborhood of a candidate edge, while every other strategy has a strictly lower payoff throughout that neighborhood. Then every distribution supported on $\{1,2\}$ is locally best-responding throughout the same neighborhood and is therefore state-robust. Robust mixing is possible, but only because the support equality is a genuine local payoff identity rather than a pointwise tie.

The full-support case gives a transparent codimension count. Let
\[
D=\dim X=\sum_{p\in P}(n_p-1).
\]
Then $\dim\mathcal A_X=D+1$.

\begin{theorem}[Full-support SRE in arbitrary finite dimension]\label{thm:fullsupport}
Suppose \(\Gamma\) is affine and \(x\in\inter(X)\) is a Nash equilibrium. Then the following statements are equivalent:
\begin{enumerate}[label=(\roman*),leftmargin=0.28in]
\item \(x\in\SRE(\Gamma)\);
\item for every population \(p\) and every pair of strategies \(i,j\in S_p\),
\[
F_{p,i}(y)=F_{p,j}(y)\qquad \forall y\in X;
\]
\item for every population \(p\) and every strategy \(i\in S_p\),
\[
 h_{p,i}(y;x)=0\qquad \forall y\in X.
\]
\end{enumerate}
Thus existence of a full-support SRE is equivalent to within-population payoff identity on \(X\), and is independent of the particular full-support state. Consequently, the affine games admitting at least one full-support SRE form a linear subspace of \(\prod_{p\in P}\mathcal A_X^{n_p}\) of codimension
\begin{equation}\label{eq:fullsupport-codim}
\sum_{p\in P}(n_p-1)(D+1).
\end{equation}
\end{theorem}

The theorem is deliberately stated for arbitrary finite-strategy population games. It shows that full-support robustness cannot be protected by a special cancellation at one mixture; the restriction is a high-codimension linear one in every finite dimension.

For the standard one-population matrix representation, Theorem~\ref{thm:fullsupport} gives a row-additive form.

\begin{corollary}[One-population matrix representation]\label{cor:onepopgeneric}
Consider a one-population affine game of unit mass with \(n\ge2\) strategies,
\[
F(y)=Ay+b,
\qquad X=\{y\in\R_+^n:\one^\top y=1\}.
\]
Let \(x\in\inter(X)\) be a fully mixed Nash equilibrium. Then \(x\in\SRE(\Gamma)\) if and only if the following equivalent conditions hold:
\begin{enumerate}[label=(\roman*),leftmargin=0.28in]
\item \(F_i(y)=F_j(y)\) for every \(i,j\) and every \(y\in X\);
\item the payoff matrix is row-additive, \(A_{ij}=u_i+r_j\) for some \(u,r\in\R^n\), and \(u_i+b_i\) is constant in \(i\);
\item there exist \(c\in\R\) and \(v\in\R^n\) such that \(F_i(y)=c+v\cdot y\) for every strategy \(i\) and every state \(y\in X\).
\end{enumerate}
The set of affine coefficients \((A,b)\in\R^{n\times n}\times\R^n\) admitting a fully mixed SRE is a linear subspace of codimension \(n(n-1)\).
\end{corollary}

The count in Corollary~\ref{cor:onepopgeneric} is taken in the displayed coefficient space $(A,b)$. The row-additive restrictions are invariant to alternative coefficient representations of the same affine functions on the simplex, so this coefficient-space statement is consistent with the quotient-space formulation above.

The codimension count has a direct interpretation. Full-support state-robustness requires not just equality of payoffs at the equilibrium point, but local equality of all pure payoff functions on the simplex. Outside the exceptional subspace, any full-support Nash equilibrium is exposed by a pure deviation. Hawk--Dove and rock--paper--scissors are therefore examples of a general full-support exposure theorem, not special pathologies of low-dimensional games.

\subsection{A smooth extension}\label{subsec:smooth}

The affine results give the finite-dimensional characterization used in the paper. For continuously differentiable but nonlinear games, the following auxiliary observation remains useful: first-order exposure is sufficient, though not necessary, because higher-order terms can create exposure when the first derivative is zero.

\begin{proposition}[First-order sufficient condition]\label{prop:smooth}
Suppose $F$ admits a continuously differentiable extension to an open neighborhood of $X$ in $\R^n$. More generally, the argument only requires that the restriction of $F$ to $\operatorname{aff}(X)$ be continuously differentiable. Fix $x\in X$, $p\in P$, and $i\in S_p$. If either $h_{p,i}(x;x)>0$, or $h_{p,i}(x;x)=0$ and there exists $d\in\ri(T_xX)$ such that
\[
D_y h_{p,i}(x;x)[d]>0,
\]
then $e_{p,i}$ structurally exposes $x$.
\end{proposition}

\section{Examples and Applications}\label{sec:applications}

\subsection{Hawk--Dove}\label{subsec:hawkdove}

\begin{example}[Hawk--Dove]\label{ex:hawkdove}
Consider a single population of unit mass with strategies $H$ and $D$. Let $x_H$ denote the mass on Hawk. With value $V>0$ and conflict cost $C>V$, random matching gives
\begin{equation}\label{eq:hd-payoffs}
F_H(x)=V-\frac{V+C}{2}x_H,
\qquad
F_D(x)=\frac{V}{2}(1-x_H).
\end{equation}
Thus
\[
F_H(x)-F_D(x)=\frac{V-Cx_H}{2}.
\]
The unique Nash equilibrium is the mixed state $x_H^*=V/C$. Since the game has one population, Observation~\ref{obs:csnb-ne} also identifies it as the unique candidate-state no-blocking prediction.

At $x^*$, Hawk and Dove earn equal payoffs. Set
\[
y_H^k=\frac VC+\frac{1-\frac VC}{k+1}.
\]
Then $y_H^k\in(V/C,1)$ for every $k$, $y_H^k\to V/C$, and Dove is strictly better than Hawk along this interior perturbation. The pure Dove gap relative to the candidate mixture is
\[
F_D(y^k)-x_H^*F_H(y^k)-(1-x_H^*)F_D(y^k)
=x_H^*\bigl(F_D(y^k)-F_H(y^k)\bigr)>0.
\]
Hence the mixed state is structurally exposed. Therefore
\[
\SRE(\Gamma)=\varnothing.
\]
The conclusion does not contradict evolutionary stability of the mixed Hawk--Dove state. Evolutionary stability studies mutant performance under a particular population perturbation. State-robustness asks whether the candidate aggregate distribution remains a best response at nearby aggregate states.

\end{example}

\subsection{Weak boundary robustness}\label{subsec:boundary}

\begin{example}[A weak boundary equilibrium that survives]\label{ex:boundary}
Consider a one-population, two-strategy game of unit mass with
\[
F_1(y)=0,\qquad F_2(y)=-y_2.
\]
At the pure state $e_1=(1,0)$, the two pure payoffs are tied: $F_1(e_1)=F_2(e_1)=0$. Hence $e_1$ is a Nash equilibrium but not a strict Nash equilibrium. Nevertheless, for every interior state $y$ with $y_2>0$, strategy 1 strictly dominates strategy 2. Therefore the candidate distribution $e_1$ remains a best response throughout every sufficiently small interior neighborhood of $e_1$, and Theorem~\ref{thm:localbr} gives $e_1\in\SRE(\Gamma)$.

The tangent certificate gives the same conclusion. At $e_1$, the tied gap of strategy 2 relative to the candidate is $h_2(y;e_1)=-y_2$. Every feasible tangent direction $d\in T_{e_1}X$ satisfies $d_2\ge0$, so $D_y h_2(e_1;e_1)[d]=-d_2\le0$. Thus the zero gap is protected by the boundary normal cone. This example separates state-robustness from strictness: SRE removes locally exposed equality, not every weak equilibrium.
\end{example}

\subsection{Payoff-identity mixing}\label{subsec:identity-example}

\begin{example}[A mixed SRE sustained by payoff identity]\label{ex:identity-mixing}
Consider a one-population unit-mass game with three strategies. Let
\[
F_1(y)=F_2(y)=0,\qquad F_3(y)=-1-y_3.
\]
Every state $x$ with $x_3=0$ is a Nash equilibrium. It is also state-robust: for every nearby evaluation state $y$, strategies 1 and 2 have identical payoff and strategy 3 is strictly worse. Hence any mixture over the institutionally equivalent strategies 1 and 2 remains a best response throughout a neighborhood. This example is deliberately degenerate, but not vacuous. It represents settings in which two labels correspond to payoff-identical compliance methods, standards, or contractual implementations, while an outside option is locally dominated. The support-equality theorem says that this type of local payoff identity is exactly what robust mixing requires.
\end{example}

\subsection{Multi-strategy coordination}\label{subsec:coordination}

Consider one population of unit mass with strategies $1,\ldots,n$ and payoffs
\begin{equation}\label{eq:coordination}
F_i(x)=x_i,
\qquad i=1,\ldots,n.
\end{equation}

\begin{proposition}[Coordination selects pure anchors]\label{prop:coordination}
In the coordination game~\eqref{eq:coordination},
\[
\SRE(\Gamma)=\{e_1,\ldots,e_n\}.
\]
\end{proposition}

The Nash set consists of the uniform distributions over non-empty subsets of strategies. The pure states are strict pure Nash equilibria and survive. Every mixed uniform state is exposed by moving mass locally toward one strategy in its support and away from another.

\subsection{Rock--paper--scissors}\label{subsec:rps}

Consider the standard rock--paper--scissors game with payoff matrix
\[
A=\begin{pmatrix}
0&-1&1\\
1&0&-1\\
-1&1&0
\end{pmatrix},
\qquad F(x)=Ax.
\]

\begin{proposition}[Cyclic exposure]\label{prop:rps}
In rock--paper--scissors,
\[
\SRE(\Gamma)=\varnothing.
\]
\end{proposition}

The unique Nash equilibrium is $x^*=(1/3,1/3,1/3)$. At $x^*$ all pure payoffs are tied. A perturbation that increases Scissors and decreases Paper makes Rock strictly profitable relative to the candidate mixture, so the unique Nash equilibrium is exposed.

\subsection{Binary symmetric diagnostic}\label{subsec:binary}

A binary symmetric population game has one population, strategies $1$ and $2$, unit mass, and payoff matrix $(m_{ij})_{i,j=1}^2$. Let $x_2$ be the mass on strategy $2$. Then
\[
F_1(x)=m_{11}(1-x_2)+m_{12}x_2,
\qquad
F_2(x)=m_{21}(1-x_2)+m_{22}x_2.
\]

\begin{corollary}[Binary diagnostic]\label{cor:binary}
Let $\Gamma$ be a generic binary symmetric population game with $m_{11}\ne m_{21}$ and $m_{22}\ne m_{12}$. Then $\SRE(\Gamma)\ne\varnothing$ if and only if the game admits a strict pure Nash equilibrium. More specifically:
\begin{enumerate}[label=(\roman*),leftmargin=0.28in]
\item If strategy $1$ strictly dominates, then $\SRE(\Gamma)=\{(1,0)\}$.
\item If strategy $2$ strictly dominates, then $\SRE(\Gamma)=\{(0,1)\}$.
\item If the game is a coordination game, then $\SRE(\Gamma)=\{(1,0),(0,1)\}$.
\item If the game is an anti-coordination game, then $\SRE(\Gamma)=\varnothing$.
\end{enumerate}
\end{corollary}

The binary classification is the smallest sign diagram for the general decomposition. Dominance and coordination have strict pure anchors. Anti-coordination has only an interior equality, and that equality is locally exposed.

The diagnostic also explains why state-robustness agrees with the usual intuition in dominance and coordination games but disagrees with evolutionary stability in anti-coordination games. In dominance games, the dominant strategy remains optimal at every nearby state. In coordination games, each pure convention is locally self-confirming. In anti-coordination games, the mixed state balances the two pure incentives exactly; moving the aggregate state to either side of the balance creates a strict pure improvement. Thus the difference is not driven by dimensionality or by an unusual payoff specification. It is already present in the smallest nontrivial population game.

\subsection{Two-sided platform adoption}\label{subsec:platform}

This subsection studies a stripped-down two-sided adoption game inspired by platform competition and two-sided-market models \citep{rochet2003platform,caillaud2003chicken,armstrong2006competition,weyl2010price}. There are two unit-mass populations: buyers ($p=1$) and sellers ($p=2$). Both choose between platforms $A$ and $B$. Let $x_{p,A}$ be the mass of side $p$ on platform $A$. Payoffs are
\begin{align}
F_{1,A}(x)&=\alpha x_{2,A}, & F_{1,B}(x)&=\beta x_{2,B}, \label{eq:platform-buyers}\\
F_{2,A}(x)&=\gamma x_{1,A}-\phi, & F_{2,B}(x)&=\delta x_{1,B}, \label{eq:platform-sellers}
\end{align}
where $\alpha,\beta,\gamma,\delta>0$ and $0\le\phi<\gamma$.

\begin{proposition}[Platform tipping]\label{prop:platform}
In the platform adoption game~\eqref{eq:platform-buyers}--\eqref{eq:platform-sellers}, the Nash set is
\[
\NE(\Gamma)=\{x^A,x^B,x^\circ\},
\]
where
\[
x^A=((1,0),(1,0)),\qquad x^B=((0,1),(0,1)),
\]
and
\[
x^\circ=\left(\left(\frac{\delta+\phi}{\gamma+\delta},\frac{\gamma-\phi}{\gamma+\delta}\right),
\left(\frac{\beta}{\alpha+\beta},\frac{\alpha}{\alpha+\beta}\right)\right).
\]
Moreover,
\[
\SRE(\Gamma)=\{x^A,x^B\}.
\]
\end{proposition}

The two tipping states are strict pure Nash equilibria. The interior state $x^\circ$ is sustained by simultaneous indifference of buyers and sellers. A small increase in buyer adoption of platform $A$ makes the seller-side pure proposal to $A$ strictly profitable. A small increase in seller adoption of $A$ makes the buyer-side pure proposal to $A$ strictly profitable. Thus the mixed adoption configuration is exposed, while both tipping states survive.

The linear-program certificate gives this rejection a direct economic interpretation. At $x^\circ$, the seller-side gap for a pure move to platform $A$ has derivative
\[
D_yh_{2,A}(x^\circ;x^\circ)[d]=x^\circ_{2,B}(\gamma+\delta)d_{1,A}.
\]
The normalized tangent program can therefore choose a feasible inward direction with $d_{1,A}>0$, for example by moving buyer mass locally from $B$ to $A$, to obtain a positive value. Economically, the exposing direction says that if buyer adoption of $A$ is locally higher than the reported interior state, then sellers strictly prefer the pure $A$ proposal to the candidate mixture. Symmetrically, the buyer-side $A$ gap has derivative $x^\circ_{1,B}(\alpha+\beta)d_{2,A}$, so a local increase in seller adoption of $A$ exposes the buyer-side pure move to $A$. The finite test therefore returns not only rejection, but also the side, platform, and aggregate-state direction responsible for the rejection.

The proposition should be read as a selection result within a deliberately spare adoption game, not as a full model of platform pricing. There are no fees, no multi-homing decisions, and no platform-side strategic variables. The point of the example is instead to isolate the role of cross-side network externalities. The interior state balances the two sides' incentives exactly. Once one side is perturbed even slightly toward a platform, the other side strictly wants to follow. State-robustness therefore recovers the set of tipping predictions from a static local test rather than from an assumed adjustment dynamic.

The refinement is intentionally silent between $x^A$ and $x^B$. If both tipping states are strict, both are locally robust. A welfare comparison, price competition, entry history, or risk-dominance criterion may select between them, but that selection uses information outside the state-robustness requirement. This limited silence is a feature of the concept: it removes locally exposed indifference, not all multiplicity.

\subsection{Standards adoption with intrinsic quality}\label{subsec:standards}

Consider one population of unit mass choosing among $n\ge2$ standards, in the spirit of models of network externalities and technological lock-in \citep{katz1985network,david1985clio,arthur1989competing}. Standard $i$ has intrinsic quality $q_i\in\R$ and same-side network effect $\lambda>0$:
\begin{equation}\label{eq:standards}
F_i(x)=q_i+\lambda x_i,
\qquad i=1,\ldots,n.
\end{equation}
Let $q^*=\max_iq_i$.

\begin{proposition}[Quality and lock-in]\label{prop:standards}
In the standards-adoption game~\eqref{eq:standards},
\[
\SRE(\Gamma)=\{e_i:q_i+\lambda>q^*\}.
\]
Every mixed Nash equilibrium is structurally exposed. If the highest-quality standard is unique, say $i^*$, and
\[
\lambda<q^*-\max_{j\ne i^*}q_j,
\]
then $e_{i^*}$ is the unique state-robust equilibrium. If
\[
\lambda>q^*-\min_iq_i,
\]
then every pure adoption state is state-robust.
\end{proposition}

The formula separates quality from network effects. A lower-quality standard survives only if its own network advantage strictly compensates its quality disadvantage. Boundary cases with $q_i+\lambda=q^*$ are not robust: the superior-quality proposal is tied at the candidate and exposed by moving an infinitesimal amount of mass toward it.

The comparative statics have the usual lock-in interpretation. When network effects are weak relative to the quality gap, the highest-quality standard is the only robust outcome. When network effects are strong, even inferior standards can be robust once they have captured the entire population, because the installed base creates a strict local advantage. Intermediate values of $\lambda$ generate multiple robust pure states. State-robustness therefore distinguishes two sources of multiplicity: robust multiplicity among strict installed-base outcomes, and fragile multiplicity among mixed adoption equilibria supported by exact equality.

The proposition also illustrates why the boundary case uses a strict inequality. If $q_i+\lambda=q^*$ for a lower-quality standard $i$, the pure state $e_i$ is a Nash equilibrium but not a robust one. A tiny interior presence of the higher-quality standard breaks the tie in its favor. Thus equality between quality disadvantage and network advantage is itself a knife edge.

\Needspace{14\baselineskip}
\section{Robustness, Related Concepts, and Scope}\label{sec:discussion}

\subsection{Payoff robustness of strict pure states}\label{subsec:payoffrobust}

State-robustness perturbs the aggregate state, not the payoff function. Strict pure equilibria nevertheless enjoy the familiar payoff robustness property.

\Needspace{8\baselineskip}
\begin{proposition}[Payoff robustness of strict pure SRE]\label{prop:payoffrobust}
Let $x^*$ be a strict pure Nash equilibrium of $\Gamma$. There exists $\eta>0$ such that, for every continuous payoff function $\widetilde F$ with
\[
\sup_{x\in X}\|F(x)-\widetilde F(x)\|_\infty<\eta,
\]
the same state $x^*$ is a state-robust equilibrium of the perturbed game $\widetilde\Gamma$ obtained by replacing $F$ with $\widetilde F$.
\end{proposition}

The proposition is a consistency check. The refinement preserves the outcomes that ordinary strictness already protects. Its force is concentrated on weak equilibria whose optimality depends on exact local indifference.

\subsection{Comparison with evolutionary stability}\label{subsec:ess}

Evolutionary stability and state-robustness agree on strict pure states but diverge on interior anti-coordination states. ESS is used here in the standard one-population symmetric-game sense.

\begin{proposition}[ESS and SRE in generic binary games]\label{prop:ess}
In every generic binary symmetric population game:
\begin{enumerate}[label=(\roman*),leftmargin=0.28in]
\item every strict pure Nash equilibrium is both evolutionarily stable and state-robust;
\item in anti-coordination games, the unique mixed Nash equilibrium is evolutionarily stable but not state-robust.
\end{enumerate}
\end{proposition}

The divergence reflects different robustness questions. ESS asks whether a resident distribution defeats mutants under a specified evolutionary comparison. State-robustness asks whether the candidate distribution remains a best response when the aggregate state used to evaluate payoffs is slightly misspecified.

\subsection{Relation to other robustness concepts}\label{subsec:otherrefinements}

SRE is easiest to place by specifying what varies and what is held fixed. The payoff map, information environment, and reported prescription are fixed. Only the aggregate state at which payoff comparisons are evaluated is perturbed. This separates SRE from several neighboring concepts.

First, SRE is not strictness in disguise. Strict pure Nash equilibria survive by Corollary~\ref{cor:strictpure}, but weak boundary equilibria may also survive when the feasible tangent cone protects the candidate from nearby strict improvement. Conversely, mixed equilibria that are perfectly legitimate Nash equilibria may fail SRE when their indifference is locally exposed. Thus the refinement removes exposed payoff equalities, not all weak equilibria.

Second, SRE is distinct from perfection, sequential equilibrium, and strategic stability \citep{selten1975reexamination,kreps1982sequential,kohlberg1986strategic,vandamme1991stability,mailath1997proper}. Those refinements perturb strategies, beliefs, trembles, or nearby strategic forms and often study stable sets of equilibria. SRE does not construct a stable set, an assessment, or a tremble system. The strategic environment is fixed, and the local test is whether the reported aggregate prescription remains a best response under nearby payoff-evaluation states.

Third, SRE differs from persistent and essential equilibrium concepts. Essential-equilibrium arguments ask whether equilibria or equilibrium components survive small perturbations of the game \citep{wujiang1962essential}. Persistent equilibrium requires membership in a persistent retract, a set-valued strategic object designed to survive admissible perturbations of opponents' strategies \citep{kalai1984persistent}. Robustness of refinements similarly studies how refinement requirements behave under perturbed games or strategic forms \citep{fudenberg1988robustness}. SRE instead holds the game and the reported prescription fixed. It may be empty, and emptiness is a diagnostic statement that no reported Nash prediction is locally valid under aggregate-state misspecification.

Fourth, SRE differs from large-game, continuous-game, and robust aggregative concepts. \citet{kalai2004large} studies large semi-anonymous games; \citet{chen2022robust} combine aggregate robustness with admissibility and ex post robust perfection in continuum-player games; \citet{lotidis2025robust} study equilibria invariant to small payoff- or gradient-field perturbations; and robust aggregative or ambiguity-set formulations robustify agents' optimization against uncertainty sets in opponents' behavior or aggregate states \citep{lanzetti2025strategically,feik2026strategically}. SRE instead fixes the payoff map and the reported prescription. It may be empty, does not perturb payoffs or the strategic form, and asks only whether a fixed reported population distribution remains optimal when the aggregate state used in payoff evaluation is locally misspecified. The tangent- and normal-cone tests characterize exactly that reported-state property in affine population games.

Fifth, SRE differs from robustness to incomplete information and higher-order beliefs. \citet{kajii1997robustness} ask when complete-information equilibrium predictions survive small departures from common knowledge of payoffs. \citet{bergemann2013robust} study robust predictions that are valid across possible private information structures and characterize the resulting outcomes through Bayes correlated equilibrium; \citet{morris2005generalized} and \citet{bergemann2016bayes} develop related robust-equilibrium and information-structure comparisons. SRE fixes the information structure and does not vary beliefs. The robustness question concerns measurement or implementation of the aggregate state, not misspecification of players' information.

Sixth, SRE differs from evolutionary stability. ESS asks whether a resident distribution resists mutants under an evolutionary comparison \citep{maynardsmith1973logic,sandholm2010population}. SRE asks whether the reported candidate distribution remains a best response when the aggregate state used to evaluate payoffs is slightly misspecified. Proposition~\ref{prop:ess} shows that the two tests can disagree even in generic binary anti-coordination games: the mixed equilibrium is evolutionarily stable but not state-robust.

Seventh, SRE is distinct from local superiority, neighborhood invader strategies, and static stability in symmetric and population games \citep{apaloo1997neighborhood,milchtaich2020static}. Those concepts compare the performance of an incumbent strategy or population state against nearby resident or mutant states, often with an evolutionary interpretation. SRE instead fixes the reported prescription $x_p$ and asks whether it remains a best response when only the payoff-evaluation aggregate state $y$ is locally misspecified. The relevant inequality is therefore
\[
z_p\cdot F_p(y)\le x_p\cdot F_p(y)
\]
for all nearby evaluation states $y$ and all population deviations $z_p$, not a superiority comparison between neighboring resident states. This distinction is what makes boundary normal-cone protection and the finite LP diagnostic possible in multi-population product simplices.

Payoff-perturbation and purification arguments change the payoff environment or introduce payoff disturbances \citep{harsanyi1973disturbed,anderlini2001structural}; SRE keeps the structural payoff map fixed. Rationalizability is also different. It checks whether strategies can be justified by beliefs consistent with iterative reasoning \citep{bernheim1984rationalizable,pearce1984rationalizable}; it does not require an aggregate state to remain locally best-responding. In the coordination game~\eqref{eq:coordination}, mixed uniform Nash states are rationalizable, but Proposition~\ref{prop:coordination} shows that they are not state-robust.

One conceptual connection is to structural rationality in \citet{siniscalchi2022structural}, where robustness is expressed through perturbations of conditional-probability systems. The state space here is instead a product of finite simplices, and robustness is evaluated through local payoff comparisons at nearby aggregate states. The resulting object is deliberately conservative: it does not replace dynamics, learning, risk dominance, global games, welfare comparisons, or payoff perturbations \citep{harsanyi1988selection,carlsson1993global}. It identifies which Nash predictions are robust to small state misspecification and which are artifacts of exact aggregate indifference.

\subsection{How the finite test is used}\label{subsec:implementation}

The affine linear-program test suggests a simple workflow for applications. First compute the Nash set, or a candidate subset of it. Second, for each candidate state $x$, compute all pure gaps $h_{p,i}(x;x)$. Strategies with strictly negative gaps are irrelevant by continuity. Strategies with positive gaps reveal that the candidate was not Nash. Only zero gaps remain. Third, for each zero gap, solve the tangent program in \eqref{eq:lp} or check the normal-cone certificate in \eqref{eq:normal-cert}. A value $\Psi_{p,i}(x)>0$ exposes the candidate; $\Psi_{p,i}(x)=0$ protects it against that deviation.

The procedure is local in two senses. It is local in the state space because only the tangent and normal cones at the candidate matter. It is also local in the payoff map because only the derivative of the relevant payoff gap at the candidate matters in affine games. This is useful for empirical work: one can report not only whether a candidate equilibrium is exposed, but also which population, which pure strategy, and which direction in the aggregate state space expose it. The exposing direction is an interpretable diagnostic rather than a black-box rejection; in the platform example, a positive buyer-side $A$ direction in the seller-side $A$-gap program says that the mixed report is invalid when buyer adoption of $A$ is locally understated.

If a fixed reported candidate is accompanied by uncertainty about the payoff-evaluation state, the same logic applies to any local uncertainty region supplied from outside the model. The result below conditions on a fixed report $x$ and is complementary to inference in many-player anonymous games, where payoff-relevant empirical distributions are estimated from data \citep{menzel2016inference}. Let $U\subseteq X$ be a local region for the payoff-relevant aggregate state around a reported candidate $x$. Say that $x$ is $U$-valid if
\begin{equation}\label{eq:U-valid}
\sup_{y\in U}h_{p,i}(y;x)\le0\quad\text{for every }p\in P\text{ and }i\in S_p.
\end{equation}
Then the reported prescription is best-responding throughout $U$. When $U$ is compact, the supremum is a maximum. If \eqref{eq:U-valid} fails, some pair $(p,i)$ has a positive supremum; when $U$ is compact, a maximizing evaluation state also identifies the concrete deviation that the data or implementation region cannot rule out. In affine games with closed polyhedral $U$, each optimization problem in \eqref{eq:U-valid} is a linear program.

Fix any norm $\|\cdot\|$ on the finite-dimensional vector space $\operatorname{span}(X-X)$. For a deterministic local analysis, write
\[
B_X(x,r)=\{y\in X:\|y-x\|\le r\}.
\]
All balls and inclusions are relative to the feasible state space $X$; at boundary reports they do not require Euclidean directions outside $X$. The inner-ball condition below is substantive: it holds for metric balls such as $U_m=B_X(x,r_m)$, but not for arbitrary shrinking sets.

\begin{proposition}[Shrinking uncertainty-region diagnostic]\label{prop:uncertainty}
Fix a continuous finite-strategy population game and a fixed reported state $x\in X$. Let $(U_m)$ be compact subsets of $X$ such that, for each $m$, there exists $\rho_m>0$ with
\[
B_X(x,\rho_m)\subseteq U_m,
\]
and there exists $\bar\rho_m\downarrow0$ with
\[
U_m\subseteq B_X(x,\bar\rho_m).
\]
Then $x\in\SRE(\Gamma)$ if and only if $x$ is $U_m$-valid for all sufficiently large $m$. If $\Gamma$ is affine and each $U_m$ is polyhedral, each $U_m$-validity check is a finite collection of linear programs.
\end{proposition}

This is a deterministic fixed-report statement. If the report itself is a random estimator $\widehat x_m$, an external consistency layer is needed; if $U_m$ is interpreted as a confidence region, coverage alone does not imply the inner-ball and shrinking conditions above.

\begin{corollary}[Asymptotic uncertainty-region implication]\label{cor:asymp-uncertainty}
Let $(U_m)$ be random compact subsets of $X$, and assume that the validity indicators below are measurable. Suppose there exist random radii $\rho_m>0$ and $\bar\rho_m\ge0$ such that $\bar\rho_m\to0$ in probability and
\[
\Pr\bigl(B_X(x,\rho_m)\subseteq U_m\subseteq B_X(x,\bar\rho_m)\bigr)\to1.
\]
Then
\[
\mathbf 1\{x\text{ is }U_m\text{-valid}\}\ \longrightarrow\ \mathbf 1\{x\in\SRE(\Gamma)\}
\quad\text{in probability}.
\]
If, in addition, $\liminf_m\Pr(\theta_m\in U_m)\ge1-\alpha$, where $\theta_m$ denotes the true payoff-evaluation aggregate state in the $m$th environment, then under $x\in\SRE(\Gamma)$,
\[
\liminf_{m\to\infty}\Pr\bigl(x_p\in\BR_p(\theta_m)\text{ for every }p\in P\bigr)\ge 1-\alpha.
\]
\end{corollary}

The proposition and corollary formalize the uncertainty-region diagnostic for a fixed reported state. Regular shrinking regions eventually accept exactly the reported-state robust candidates and reject exactly the candidates with a nearby strict profitable pure deviation. Rejection remains interpretable: it comes with a population, a pure strategy, and, for compact regions, a payoff-evaluation state that violates the reported best-response inequalities.

The test also avoids a common ambiguity in equilibrium refinements. It does not require choosing a metric over perturbed games, a distribution over trembles, or a dynamic adjustment rule. The only primitives are the payoff function and the feasible state space. This makes the refinement modest but transparent. If a candidate fails the test, the analyst knows exactly what additional structure would be needed to defend it: a reason why the exposing state perturbations should not be considered relevant, or a dynamic/institutional mechanism that selects the candidate despite local best-response failure.

\subsection{Scope}\label{subsec:limitations}

The refinement is intentionally not a universal selection criterion. It preserves all strict pure equilibria, so it does not resolve multiplicity among robust conventions. In platform adoption, both tipping states survive. In coordination games, all pure conventions survive. This is appropriate: state-robustness removes predictions that depend on locally exposed equality, but it does not rank strict local anchors. Welfare, history, risk dominance, stochastic stability, bargaining, or institutional design may be needed for that second task.

The finite-strategy assumption is also substantive. The pure-deviation reduction relies on the fact that a mixed improvement contains a pure improvement and that only finitely many pure strategies must be checked. With compact continuous strategy spaces, an analogous theorem would require additional regularity ensuring that improving pure strategies do not drift without a convergent subsequence. The tangent-cone geometry would also move from finite-dimensional polyhedral cones to infinite-dimensional tangent objects. Those extensions are natural but outside the scope of the present paper.

Finally, the paper fixes candidate-state payoff evaluation. This convention is common and tractable, but it is not the only possible cooperative benchmark. If a coalition's proposed reallocation changes the state used to evaluate payoffs, then coalitional externalities re-enter and non-singleton coalitions can matter. Such a post-deviation theory would be genuinely different. It would require new definitions, new existence questions, and different geometric tests because the proposal would enter the payoff state itself. The contribution here is to give a finite-dimensional and internally consistent robustness theory for the candidate-state benchmark.


\section{Conclusion}\label{sec:conclusion}

This paper develops state-robust equilibrium as a reported-state diagnostic for Nash predictions in finite-strategy population games. The candidate distribution and payoff map are held fixed; only the payoff-evaluation state varies. This distinction turns exact pointwise indifference into a local robustness question.

The candidate-state no-blocking observation fixes the benchmark: when coalition proposals do not enter the state at which payoffs are evaluated, blocking collapses to singleton best-response checks and the no-blocking set is Nash. The substantive refinement is the subsequent reported-state robustness test.

The main continuous-game result gives four equivalent readings of SRE: local best-response invariance, no structural exposure, persistence of candidate best responses along every vanishing sequence of interior aggregate-state errors, and absence of nearby pure strict-improvement regions. The maximality corollary says that, once arbitrary vanishing state error is imposed as the validity criterion, SRE is the largest prediction correspondence that can be retained.

For affine games, zero payoff gaps are exposed exactly when an inward tangent direction raises the gap; non-exposure is certified by membership of the gap derivative in the normal cone. The resulting linear programs identify whether a candidate fails and, if so, the exposing population, strategy, and aggregate-state direction. The exposed-indifference decomposition separates Nash equilibria into locally protected predictions and predictions supported by exposed indifference.

The generic implications are intentionally sharp. Mixed SRE require local payoff identity on their supports, and affine mixed SRE require global payoff identity on the state space. Hence outside payoff-identity subspaces every affine SRE is pure; after excluding payoff-tie hyperplanes at pure states, SRE coincides with strict pure Nash equilibria. This is a consequence, not a definition: robust mixing survives under genuine payoff identity, and weak boundary equilibria can survive when the feasible tangent cone protects them.

The uncertainty-region diagnostic makes the concept operational. A reported state accompanied by a local region of possible payoff-evaluation states can be checked by the same inequalities; in affine games with polyhedral regions this check is finite. Regular shrinking regions converge exactly to SRE, and statistical confidence regions can be used when an external procedure supplies the stated shrinking, inner-ball, measurability, and coverage conditions. If the state-robust set is empty, the game has no static aggregate prediction whose best-response content survives small aggregate-state misspecification. Prediction may still be possible, but it must then use additional structure--dynamics, history, learning, institutions, risk dominance, welfare criteria, or a genuinely post-deviation coalitional concept.

Natural extensions include continuous strategy spaces, sequential population games, and empirical implementations that condition on reported aggregate states and combine them with statistically justified local regions satisfying explicit regularity and coverage conditions. The common theme is to distinguish local best-response persistence from exact pointwise indifference.

\appendix
\section{Proofs}\label{sec:proofs}

\subsection{Proof of Lemma~\ref{lem:singleton}}

If a singleton exposes $x$, then some coalition exposes $x$. Conversely, suppose $S$ exposes $x$ at proposal $z_S$ along perturbation $(y^k)$. Pick any $p\in S$. Then
\[
z_p\cdot F_p(y^k)>x_p\cdot F_p(y^k)
\]
for all sufficiently large $k$. Thus the singleton population $\{p\}$ exposes $x$ at proposal $z_p$ along the same perturbation.

\subsection{Proof of Lemma~\ref{lem:interior-tangent}}

For each population $q$,
\[
T_{x_q}X_q=\left\{d_q\in\R^{n_q}:\sum_{j\in S_q}d_{q,j}=0,\ d_{q,j}\ge0\text{ whenever }x_{q,j}=0\right\}.
\]
Its relative interior is obtained by replacing the weak inequalities at inactive strategies by strict inequalities. If $y\in\inter(X)$ and $x_{q,j}=0$, then $(y_q-x_q)_j=y_{q,j}>0$, while the mass constraints give $\sum_j(y_q-x_q)_j=0$; hence $y-x\in\ri(T_xX)$. Conversely, if $d\in\ri(T_xX)$, then $d_{q,j}>0$ whenever $x_{q,j}=0$. For coordinates with $x_{q,j}>0$, choose $t>0$ small enough that $x_{q,j}+td_{q,j}>0$. The mass constraints are preserved by $d\in T_xX$. Thus $x+td\in\inter(X)$ for all sufficiently small $t>0$.

\subsection{Proof of Lemma~\ref{lem:representative}}

Let $a$ and $a'$ represent the same affine derivative on $X$. Then $(a-a')\cdot(y-y')=0$ for all $y,y'\in X$, so $a-a'\in L^\perp$. Since $T_xX\subseteq L$ for every $x\in X$, we have $(a-a')\cdot d=0$ for every $d\in T_xX$. Therefore tangent directional derivatives are identical under the two representatives. Moreover, $a\cdot d\le0$ for all $d\in T_xX$ if and only if $a'\cdot d\le0$ for all $d\in T_xX$, so membership in $N_X(x)$ is also representation-invariant.

\subsection{Proof of Theorem~\ref{thm:localbr}}

First note that, for any $y\in X$,
\[
x_p\in\BR_p(y)
\quad\Longleftrightarrow\quad
h_{p,i}(y;x)\le0\text{ for every }i\in S_p.
\]
Indeed, the maximum of $z_p\cdot F_p(y)$ over $X_p$ is attained at a pure strategy.

The equivalence between (ii) and (iv) follows from Proposition~\ref{prop:pure} and Proposition~\ref{prop:closure}. A state is structurally exposed if and only if some pure strict-improvement region has $x$ in its closure.

To prove (iv) implies (i), observe that for each pair $(p,i)$ the condition
\[
x\notin\cl\{y\in\inter(X):h_{p,i}(y;x)>0\}
\]
implies that there is a relative neighborhood $V_{p,i}$ of $x$ in $X$ such that $h_{p,i}(y;x)\le0$ for every $y\in V_{p,i}\cap\inter(X)$. Since there are finitely many pairs $(p,i)$, the intersection $V=\cap_{p,i}V_{p,i}$ is a relative neighborhood of $x$. On $V\cap\inter(X)$, all pure gaps are non-positive, so $x_p\in\BR_p(y)$ for all $p$. Hence $x\in\SRE(\Gamma)$.

Statement (i) implies (iii) because every perturbation of $x$ eventually enters the neighborhood in Definition~\ref{def:sre}.

To prove (iii) implies (iv), suppose (iv) fails for some $(p,i)$. Then there are interior states $y^k\to x$ with $h_{p,i}(y^k;x)>0$. Along this perturbation, $x_p\notin\BR_p(y^k)$ for every $k$, contradicting (iii).

Finally, if $x\in\SRE(\Gamma)$, then by Definition~\ref{def:sre} there is a relative neighborhood $V$ of $x$ such that $h_{p,i}(y;x)\le0$ for every $p,i$ and every $y\in V\cap\inter(X)$. Choose $\omega\in\inter(X)$ and set $y^k=(1-1/k)x+(1/k)\omega$. For all large $k$, $y^k\in V\cap\inter(X)$, so $h_{p,i}(y^k;x)\le0$ for every $p,i$. Continuity gives $h_{p,i}(x;x)\le0$ for every $p,i$, which is equivalent to $x_p\in\BR_p(x)$ for every $p$. Thus $x\in\NE(\Gamma)$.

\subsection{Proof of Theorem~\ref{thm:tangent}}

Because $h_{p,i}(\cdot;x)$ is affine, the set $\{y\in\inter(X):h_{p,i}(y;x)>0\}$ is the intersection of $\inter(X)$ with an open half-space.

If $h_{p,i}(x;x)>0$, continuity implies that $h_{p,i}(y;x)>0$ for all interior $y$ sufficiently close to $x$, so exposure holds. If $h_{p,i}(x;x)<0$, continuity implies that the gap remains negative near $x$, so exposure fails.

It remains to consider $h_{p,i}(x;x)=0$. Suppose first that $e_{p,i}$ exposes $x$. Then there are $y^k\in\inter(X)$ with $y^k\to x$ and $h_{p,i}(y^k;x)>0$. For some $k$, $y^k\ne x$; set $d=y^k-x$. By Lemma~\ref{lem:interior-tangent}, $d\in\ri(T_xX)$. Affineness gives
\[
D_y h_{p,i}(x;x)[d]=h_{p,i}(y^k;x)-h_{p,i}(x;x)>0.
\]

Conversely, suppose there exists $d\in\ri(T_xX)$ with $D_y h_{p,i}(x;x)[d]>0$. By Lemma~\ref{lem:interior-tangent}, $x+td\in\inter(X)$ for all sufficiently small $t>0$. Affineness and $h_{p,i}(x;x)=0$ give
\[
h_{p,i}(x+td;x)=tD_y h_{p,i}(x;x)[d]>0.
\]
Thus $x+d/k$ is eventually a witnessing perturbation.

\subsection{Proof of Theorem~\ref{thm:knife}}

Let $x\in\NE(\Gamma)\setminus\SRE(\Gamma)$. By Proposition~\ref{prop:pure}, some pure deviation $e_{p,i}$ exposes $x$. Since $x$ is Nash, $h_{p,i}(x;x)\le0$. Theorem~\ref{thm:tangent} then implies $h_{p,i}(x;x)=0$ and the existence of a direction $d\in\ri(T_xX)$ with $D_y h_{p,i}(x;x)[d]>0$. Hence $x\in K_{p,i}(\Gamma)$.

Conversely, if $x\in K_{p,i}(\Gamma)$, Theorem~\ref{thm:tangent} implies that $e_{p,i}$ exposes $x$. Since $x\in\NE(\Gamma)$ by definition of $K_{p,i}$, $x\in\NE(\Gamma)\setminus\SRE(\Gamma)$.

\subsection{Proof of Theorem~\ref{thm:support-equality}}

By Theorem~\ref{thm:localbr}, there is a relative neighborhood $V$ of $x$ in $X$ such that $x_p\in\BR_p(y)$ for every $p$ and every $y\in V\cap\inter(X)$. Fix $p$ and $y\in V\cap\inter(X)$. Since $x_p$ is an optimal distribution on the simplex $X_p$, every pure strategy in $\supp(x_p)$ must attain the maximum payoff at $y$; otherwise moving mass away from a strictly inferior support strategy to a maximizer would increase $x_p\cdot F_p(y)$. Hence $F_{p,i}(y)=F_{p,j}(y)$ for all $i,j\in\supp(x_p)$ and all $y\in V\cap\inter(X)$. The set $V\cap\inter(X)$ is dense in $V$, and payoffs are continuous, so the equality extends to every $y\in V$.

If $\Gamma$ is affine, the difference $F_{p,i}-F_{p,j}$ is an affine function on $X$. An affine function that vanishes on a nonempty relatively open subset of $X$ vanishes on the affine hull of $X$, and hence on all of $X$. This proves the global statement.

\subsection{Proof of Theorem~\ref{thm:fullsupport}}

Because \(x\in\inter(X)\) is Nash, every pure strategy of every population is a best response at \(x\). Hence \(h_{p,i}(x;x)=0\) for all \(p\) and \(i\).

First prove (i) \(\Rightarrow\) (ii). Since \(x\) has full support, Theorem~\ref{thm:support-equality} implies that, for every population \(p\), all pure payoff functions \(F_{p,i}\) are identical on \(X\). This is (ii).

Statement (ii) implies (iii) because every pure payoff equals the candidate average at every state. Statement (iii) implies (i) because no pure deviation yields a strict gain at any state \(y\in X\); Proposition~\ref{prop:pure} then rules out structural exposure.

It remains only to note that (iii) implies (ii). Since \(x_p\) has full support and total mass \(m_p\), condition (iii) says
\[
m_pF_{p,i}(y)=x_p\cdot F_p(y)
\]
for every \(i\in S_p\) and every \(y\in X\). For fixed \(y\), the weights \(x_{p,i}/m_p\) are strictly positive and sum to one. If every component equals this weighted average, no component can lie above or below it; hence all \(F_{p,i}(y)\) are equal across \(i\). This proves (ii). Conversely, if all pure payoff functions within population \(p\) are identical, then every pure payoff equals the candidate average and (iii) holds.

For the codimension count, each population \(p\) has \(n_p\) affine payoff functions in the \((D+1)\)-dimensional space \(\mathcal A_X\). Requiring these \(n_p\) functions to be identical imposes \((n_p-1)(D+1)\) independent linear restrictions. Summing across populations gives \eqref{eq:fullsupport-codim}. The resulting set is a linear subspace because equality of affine functions is linear.

\subsection{Proof of Proposition~\ref{prop:closure}}

If $z_p$ exposes $x$, then there is a perturbation $(y^k)$ of $x$ such that $y^k\in\mathcal U_p(z_p;x)$ for all sufficiently large $k$. Hence $x\in\cl\mathcal U_p(z_p;x)$.

Conversely, if $x\in\cl\mathcal U_p(z_p;x)$, then because $X$ is metrizable there exists a sequence $y^k\in\mathcal U_p(z_p;x)$ with $y^k\to x$. By definition $y^k\in\inter(X)$ and $z_p\cdot F_p(y^k)>x_p\cdot F_p(y^k)$ for every $k$. Thus $z_p$ exposes $x$.

\subsection{Proof of Proposition~\ref{prop:pure}}

The if direction is immediate. For the only-if direction, suppose $x$ is exposed by population $p$ at mixed proposal $z_p$. Let $(y^k)$ be a witnessing perturbation. Since $z_p\in X_p$ has total mass $m_p$, write
\[
z_p=\sum_{i\in S_p}\alpha_i e_{p,i},\qquad \alpha_i=\frac{z_{p,i}}{m_p}\ge0,
\qquad \sum_i\alpha_i=1.
\]
For all large $k$,
\[
\sum_i\alpha_i e_{p,i}\cdot F_p(y^k)>x_p\cdot F_p(y^k).
\]
Hence for each such $k$ there exists $i(k)\in S_p$ such that
\[
e_{p,i(k)}\cdot F_p(y^k)>x_p\cdot F_p(y^k).
\]
Since $S_p$ is finite, some $i$ occurs along an infinite subsequence. That subsequence is still a perturbation of $x$, and $e_{p,i}$ exposes $x$.

\subsection{Proof of Proposition~\ref{prop:normal}}

Let $a=D_y h_{p,i}(x;x)$ and write $T=T_xX$. By Theorem~\ref{thm:tangent}, exposure at a zero gap is equivalent to the existence of $d\in\ri(T)$ with $a\cdot d>0$. If $T=\{0\}$, then $\ri(T)=\{0\}$, no such positive direction exists, and $a\cdot d\le0$ for every $d\in T$ holds trivially. Hence the claim is immediate in this case.

Suppose now that $T\ne\{0\}$. The existence of $d\in\ri(T)$ with $a\cdot d>0$ is equivalent to the existence of $d\in T$ with $a\cdot d>0$. The nontrivial direction is standard for convex cones: choose any $r\in\ri(T)$. For every $d\in T$ and every sufficiently small $\lambda\in(0,1)$, $(1-\lambda)d+\lambda r\in\ri(T)$; if $a\cdot d>0$, $\lambda$ can be chosen small enough that $a\cdot((1-\lambda)d+\lambda r)>0$. Therefore non-exposure is equivalent to $a\cdot d\le0$ for all $d\in T$, which is exactly $a\in N_X(x)$.

\subsection{Proof of Proposition~\ref{prop:smooth}}

If $h_{p,i}(x;x)>0$, continuity gives exposure. If $h_{p,i}(x;x)=0$ and there exists $d\in\ri(T_xX)$ with $D_y h_{p,i}(x;x)[d]>0$, then $x+td\in\inter(X)$ for all sufficiently small $t>0$ and
\[
h_{p,i}(x+td;x)=tD_y h_{p,i}(x;x)[d]+o(t)>0
\]
for all sufficiently small $t>0$. Thus $x+d/k$ eventually witnesses exposure.

\subsection{Proof of Proposition~\ref{prop:coordination}}

In the game $F_i(x)=x_i$, a state is Nash if and only if all strategies in its support have maximal population share. Therefore the Nash states are exactly the uniform distributions over non-empty subsets $K\subseteq\{1,\ldots,n\}$. By Corollary~\ref{cor:strictpure}, every pure state is in $\SRE(\Gamma)$.

Let $x^K$ be the uniform distribution on a set $K$ with $|K|\ge2$. Choose distinct $i,j\in K$. If $x^K$ is interior relative to the full simplex, take direction $d=e_i-e_j$. If $x^K$ is on the boundary, choose $\eps>0$ small and define $d_i=1$, $d_j=-1-(n-|K|)\eps$, $d_\ell=\eps$ for $\ell\notin K$, and $d_\ell=0$ for $\ell\in K\setminus\{i,j\}$. Then $d\in\ri(T_{x^K}X)$. In either case, for sufficiently small $\eps$ if needed,
\[
D_y h_i(x^K;x^K)[d]=d_i-x^K\cdot d>0.
\]
Theorem~\ref{thm:tangent} implies that $e_i$ exposes $x^K$. Hence no mixed Nash state is state-robust.

\subsection{Proof of Proposition~\ref{prop:rps}}

The unique Nash equilibrium is $x^*=(1/3,1/3,1/3)$. Since the game has one population, all candidate-state no-blocking predictions are Nash by Observation~\ref{obs:csnb-ne}. At $x^*$, every pure payoff and the candidate aggregate payoff are zero. Let the pure proposal be Rock. For $0<\eps<1/3$, set $y^\eps=(1/3,1/3-\eps,1/3+\eps)$. Then
\[
F_R(y^\eps)=-y_P^\eps+y_S^\eps=2\eps>0.
\]
The column sums of $A$ are zero, so $x^*\cdot F(y^\eps)=0$. Thus Rock exposes $x^*$, and $\SRE(\Gamma)=\varnothing$.

\subsection{Proof of Proposition~\ref{prop:platform}}

Let
\[
\tau_2=\frac{\beta}{\alpha+\beta},\qquad
\tau_1=\frac{\delta+\phi}{\gamma+\delta}.
\]
Under $\alpha,\beta,\gamma,\delta>0$ and $0\le\phi<\gamma$, both thresholds lie in $(0,1)$. Buyers strictly prefer $A$ if $x_{2,A}>\tau_2$, strictly prefer $B$ if $x_{2,A}<\tau_2$, and are indifferent if $x_{2,A}=\tau_2$. Sellers strictly prefer $A$ if $x_{1,A}>\tau_1$, strictly prefer $B$ if $x_{1,A}<\tau_1$, and are indifferent if $x_{1,A}=\tau_1$. These two best-response rules have exactly three fixed points. If $x_{2,A}>\tau_2$, then buyers choose $A$, so $x_{1,A}=1>\tau_1$ and sellers choose $A$, giving $x^A$. If $x_{2,A}<\tau_2$, then buyers choose $B$, so $x_{1,A}=0<\tau_1$ and sellers choose $B$, giving $x^B$. If $x_{2,A}=\tau_2$, seller optimality rules out $x_{1,A}\ne\tau_1$, because $x_{1,A}>\tau_1$ would require $x_{2,A}=1$ and $x_{1,A}<\tau_1$ would require $x_{2,A}=0$. Hence $x_{1,A}=\tau_1$, giving $x^\circ$. Thus the Nash set is exactly $\{x^A,x^B,x^\circ\}$.

The states $x^A$ and $x^B$ are strict pure Nash equilibria. At $x^A$, buyers strictly prefer $A$ because $\alpha>0$, and sellers strictly prefer $A$ because $\gamma-\phi>0$. At $x^B$, buyers strictly prefer $B$ because $\beta>0$, and sellers strictly prefer $B$ because $\delta>0$. Corollary~\ref{cor:strictpure} implies $x^A,x^B\in\SRE(\Gamma)$.

It remains to expose $x^\circ$. Consider the seller-side pure proposal to platform $A$. At $x^\circ$ the seller is indifferent, so
\[
h_{2,A}(x^\circ;x^\circ)=F_{2,A}(x^\circ)-x_2^\circ\cdot F_2(x^\circ)=0.
\]
The directional derivative of this gap in direction $d$ is
\[
D_y h_{2,A}(x^\circ;x^\circ)[d]
=x_{2,B}^\circ(\gamma+\delta)d_{1,A},
\]
because only the buyer-side adoption share changes the seller payoff difference. Since $x^\circ$ is interior, choose $d\in\ri(T_{x^\circ}X)$ with $d_{1,A}>0$. The derivative is strictly positive, so Theorem~\ref{thm:tangent} exposes $x^\circ$. Hence $\SRE(\Gamma)=\{x^A,x^B\}$.

\subsection{Proof of Proposition~\ref{prop:standards}}

At the pure state $e_i$, strategy $i$ yields payoff $q_i+\lambda$ and every other strategy $j$ yields $q_j$. Thus $e_i$ is a strict pure Nash equilibrium if and only if
\[
q_i+\lambda>q_j\quad\forall j\ne i.
\]
This is equivalent to $q_i+\lambda>q^*$, with the convention that top-quality strategies automatically satisfy the inequality because $\lambda>0$. Corollary~\ref{cor:strictpure} gives
\[
\{e_i:q_i+\lambda>q^*\}\subseteq\SRE(\Gamma).
\]

If $q_i+\lambda=q^*$ for a non-top strategy $i$, then $e_i$ is Nash but not strict. Let $j$ be a top-quality strategy. At $e_i$, the pure proposal $j$ has zero gap. Choose $\eps>0$ and define a tangent direction by
\[
d_j=1,\qquad d_\ell=\eps\ \text{ for }\ell\notin\{i,j\},\qquad d_i=-1-(n-2)\eps.
\]
Then $d\in\ri(T_{e_i}X)$, and the directional derivative of the $j$-gap is
\[
D_y h_j(e_i;e_i)[d]=\lambda(d_j-d_i)=\lambda\bigl(2+(n-2)\eps\bigr)>0.
\]
Theorem~\ref{thm:tangent} exposes $e_i$.

Now consider any mixed Nash equilibrium $x$ with support $K$ and $|K|\ge2$. All strategies in $K$ earn a common payoff. Fix distinct $i,j\in K$. The pure gap $h_i(x;x)$ is zero. If $K=\{1,\ldots,n\}$, set $d_i=1$, $d_j=-1$, and $d_\ell=0$ otherwise. If $K$ is a proper subset, let $r=n-|K|$, choose $\eps>0$, and set $d_i=1$, $d_j=-1-r\eps$, $d_\ell=\eps$ for $\ell\notin K$, and $d_\ell=0$ for $\ell\in K\setminus\{i,j\}$. In both cases $d\in\ri(T_xX)$ for small positive $\eps$ when needed, and
\[
D_y h_i(x;x)[d]=\lambda(d_i-x\cdot d)>0,
\]
because $d_i-x\cdot d=1-x_i+x_j$ in the full-support case and $d_i-x\cdot d=1-x_i+x_j+r\eps x_j$ in the boundary case. Theorem~\ref{thm:tangent} exposes $x$. Hence no mixed Nash equilibrium is state-robust.

The comparative-static statements follow immediately from the formula $\SRE(\Gamma)=\{e_i:q_i+\lambda>q^*\}$.

\subsection{Proof of Proposition~\ref{prop:payoffrobust}}

Let $x^*$ be a strict pure Nash equilibrium and let $i_p^*$ be the pure strategy used by population $p$. Populations with only one strategy have no deviations and can be ignored. If no population has an alternative strategy, the claim is immediate. Otherwise define
\[
\Delta=\min_{p\in P:\, n_p\ge2}\min_{j\ne i_p^*}\bigl(F_{p,i_p^*}(x^*)-F_{p,j}(x^*)\bigr)>0.
\]
Choose $\eta<\Delta/4$. If $\sup_x\|F(x)-\widetilde F(x)\|_\infty<\eta$, then
\[
\widetilde F_{p,i_p^*}(x^*)-\widetilde F_{p,j}(x^*)>\Delta-2\eta>\Delta/2
\]
for every $p$ and every $j\ne i_p^*$. By continuity of $\widetilde F$, these strict inequalities hold throughout a neighborhood of $x^*$. Therefore $x_p^*$ remains the unique best response to every nearby state for every population $p$. Theorem~\ref{thm:localbr} implies $x^*\in\SRE(\widetilde\Gamma)$.

\subsection{Proof of Proposition~\ref{prop:ess}}

For part (i), let $i$ be a strict pure Nash equilibrium. Against resident $i$, strategy $i$ earns strictly more than every mutant pure strategy; by linearity it earns strictly more than every mixed mutant assigning less than full mass to $i$. Hence $i$ satisfies the ESS inequality. Corollary~\ref{cor:strictpure} gives state-robustness.

For part (ii), write $q$ for the mass on strategy 2 and
\[
\Delta(q)=F_1(q)-F_2(q)=(m_{11}-m_{21})(1-q)+(m_{12}-m_{22})q.
\]
In a generic anti-coordination game, $m_{21}>m_{11}$ and $m_{12}>m_{22}$, so $\Delta(0)<0<\Delta(1)$ and the unique mixed Nash state $q^*\in(0,1)$ satisfies $\Delta(q^*)=0$ with $\Delta'(q^*)>0$. For any mutant state $q\ne q^*$, the resident payoff against the mutant minus the mutant payoff against itself is
\[
( q-q^*)\Delta(q)=\Delta'(q^*)(q-q^*)^2>0.
\]
Thus the mixed state is evolutionarily stable under the standard symmetric-game ESS criterion. Corollary~\ref{cor:binary} shows that the state-robust set is empty in the same anti-coordination case, so the mixed equilibrium is not state-robust.


\subsection{Proof of Proposition~\ref{prop:uncertainty}}

First suppose $x\in\SRE(\Gamma)$. By Definition~\ref{def:sre}, density of $\inter(X)$ in $X$, and continuity of the pure gaps, there is a relative neighborhood $V$ of $x$ such that $h_{p,i}(y;x)\le0$ for every $p,i$ and every $y\in V$. Since $U_m\subseteq B_X(x,\bar\rho_m)$ and $\bar\rho_m\downarrow0$, we have $U_m\subseteq V$ for all sufficiently large $m$. Hence $x$ is $U_m$-valid for all sufficiently large $m$.

Conversely, suppose $x\notin\SRE(\Gamma)$. By Theorem~\ref{thm:localbr}, there exist $p$ and $i$ such that
\[
x\in\cl\{y\in\inter(X):h_{p,i}(y;x)>0\}.
\]
For each $m$, the set $B_X(x,\rho_m)$ contains a relative open ball around $x$ in $X$, so it contains some $y_m\in\inter(X)$ with $h_{p,i}(y_m;x)>0$. Because $B_X(x,\rho_m)\subseteq U_m$, condition \eqref{eq:U-valid} fails for $U_m$. Thus $x$ cannot be $U_m$-valid eventually. The final linear-program statement follows because, when $\Gamma$ is affine and $U_m$ is compact and polyhedral, each function $h_{p,i}(\cdot;x)$ is affine, the supremum in \eqref{eq:U-valid} is a maximum, and each maximization is a linear program over a polyhedron.

\subsection{Proof of Corollary~\ref{cor:maximality}}

Theorem~\ref{thm:localbr} immediately implies that $\Gamma\mapsto\SRE(\Gamma)$ satisfies Definition~\ref{def:validity}. Conversely, let $R$ be any correspondence satisfying Definition~\ref{def:validity}, and take $x\in R(\Gamma)$. If $x\notin\SRE(\Gamma)$, then statement (iii) of Theorem~\ref{thm:localbr} fails. Hence there exists a perturbation $(y^k)$ of $x$ such that, for every $K$, there are $k\ge K$ and a population $p(k)\in P$ with
\[
x_{p(k)}\notin\BR_{p(k)}(y^k).
\]
No single $K$ can therefore make all population best-response requirements hold for all $k\ge K$, contradicting Definition~\ref{def:validity} for the retained prediction $x\in R(\Gamma)$. Hence $x\in\SRE(\Gamma)$, proving $R(\Gamma)\subseteq\SRE(\Gamma)$ for every game.

\subsection{Proof of Corollary~\ref{cor:strictpure}}

Let $x^*$ be a strict pure Nash equilibrium. For each population $p$, let $i_p^*$ be the pure strategy used at $x^*$. Strictness gives
\[
F_{p,i_p^*}(x^*)>F_{p,j}(x^*)\quad\text{for every }j\ne i_p^*.
\]
By continuity, the same inequalities hold in a neighborhood of $x^*$. Hence $x_p^*$ remains the unique best response for every population throughout that neighborhood. Theorem~\ref{thm:localbr} gives $x^*\in\SRE(\Gamma)$.

\subsection{Proof of Corollary~\ref{cor:lp}}

By Proposition~\ref{prop:pure}, it suffices to check pure deviations. If $h_{p,i}(x;x)>0$, Theorem~\ref{thm:tangent} gives exposure. If $h_{p,i}(x;x)<0$, it gives non-exposure. If $h_{p,i}(x;x)=0$, Theorem~\ref{thm:tangent} says exposure occurs iff there exists $d\in\ri(T_xX)$ with positive directional derivative. As in the proof of Proposition~\ref{prop:normal}, this is equivalent to the existence of $d\in T_xX$ with positive directional derivative. The normalization $\|d\|_\infty\le1$ is without loss because the inequalities are homogeneous. Thus exposure at a zero gap occurs iff $\Psi_{p,i}(x)>0$. There are exactly $\sum_pn_p$ pure deviations to check.

\subsection{Proof of Corollary~\ref{cor:generic-purity}}

For each triple $(p,i,j)$ with $i<j$, the restriction $F_{p,i}=F_{p,j}$ on $X$ is a proper linear subspace of $\prod_{q\in P}\mathcal A_X^{n_q}$, since it imposes equality of two affine functions on $X$. There are finitely many such triples. If a game lies outside their union, Theorem~\ref{thm:support-equality} rules out any SRE in which some population assigns positive mass to two distinct strategies. Thus every SRE is pure.

Now also exclude payoff ties at pure states. A pure state is Nash if and only if every population's chosen pure strategy weakly dominates its unchosen pure strategies at that state. Under the no-tie condition, every pure Nash equilibrium is strict. Corollary~\ref{cor:strictpure} then implies that every pure Nash equilibrium is an SRE, while Theorem~\ref{thm:localbr} implies that every SRE is Nash. Hence $\SRE(\Gamma)$ equals the set of strict pure Nash equilibria.

\subsection{Proof of Corollary~\ref{cor:onepopgeneric}}

The equivalence of SRE with equality of all pure payoff functions on \(X\) follows from Theorem~\ref{thm:fullsupport}. It remains to translate this equality into the matrix representation. If \(F_i(y)=F_j(y)\) for every \(i,j\) and every \(y\in X\), then for each \(i\) the affine difference \((A_i-A_1)y+(b_i-b_1)\) vanishes on the simplex. Hence there is a scalar \(\lambda_i\) such that
\[
A_i-A_1=\lambda_i\one^\top,\qquad b_i-b_1=-\lambda_i.
\]
Setting \(u_i=\lambda_i\) and \(r_j=A_{1j}\) yields \(A_{ij}=u_i+r_j\), after absorbing the normalization \(u_1=0\) into \(r\). The residual condition \(u_i+b_i\) constant follows from \(b_i-b_1=-\lambda_i\). Conversely, if \(A_{ij}=u_i+r_j\) and \(u_i+b_i=c\) for all \(i\), then for every \(y\in X\),
\[
F_i(y)=\sum_j(u_i+r_j)y_j+b_i=u_i+b_i+r\cdot y=c+r\cdot y,
\]
which is independent of \(i\). This proves the equivalence with the common-payoff representation \(F_i(y)=c+v\cdot y\).

The codimension count can be read either from Theorem~\ref{thm:fullsupport}, where \(D=n-1\) and the quotient-space codimension is \((n-1)n\), or directly from the displayed coefficient representation. In coefficient space, row-additive matrices have dimension \(2n-1\) inside \(\R^{n\times n}\), giving codimension \((n-1)^2\), and the residual condition imposes \(n-1\) additional independent restrictions on \(b\). The total is \((n-1)^2+(n-1)=n(n-1)\).

\subsection{Proof of Corollary~\ref{cor:binary}}

The four generic sign patterns exhaust binary symmetric games. In strict-dominance cases, the unique Nash equilibrium is the corresponding strict pure state, which is state-robust by Corollary~\ref{cor:strictpure}. In coordination games, both pure states are strict pure Nash equilibria and hence state-robust. The unique interior mixed Nash equilibrium is exposed because the affine payoff difference crosses zero at that point; moving the state to one side of the crossing makes one pure strategy strictly better than the candidate mixture. In anti-coordination games, the only Nash equilibrium is the interior mixed state, and the same crossing argument exposes it. Therefore the stated SRE sets follow.

\subsection{Proof of Corollary~\ref{cor:asymp-uncertainty}}

On the event that the regularity inclusions hold, Proposition~\ref{prop:uncertainty} applies pointwise. If $x\in\SRE(\Gamma)$, the outer inclusion $U_m\subseteq B_X(x,\bar\rho_m)$ and $\bar\rho_m\to0$ in probability imply that $U_m$ is contained in the deterministic SRE neighborhood with probability tending one; hence $x$ is $U_m$-valid with probability tending one. If $x\notin\SRE(\Gamma)$, the inner inclusion $B_X(x,\rho_m)\subseteq U_m$ with $\rho_m>0$ implies, exactly as in the second half of Proposition~\ref{prop:uncertainty}, that $U_m$ is invalid on the regularity event. This proves convergence of the indicator in probability. If $x$ is $U_m$-valid and $\theta_m\in U_m$, then \eqref{eq:U-valid} gives $x_p\in\BR_p(\theta_m)$ for every $p$. Under $x\in\SRE(\Gamma)$, $U_m$-validity has probability tending one; combining this with the coverage condition gives the stated lower bound.

\begin{spacing}{1.0}

\end{spacing}

\end{document}